\documentclass[12pt,cite,epsfig,psfrag]{article}
\usepackage[utf8]{inputenc}
\usepackage{float}
\usepackage{placeins}
\usepackage{geometry}
 \usepackage{graphicx}
 \usepackage{subfig}
  \usepackage{amsmath}
  \graphicspath{{Images@hp/}}
  \usepackage[usenames, dvipsnames]{color}
\usepackage{wrapfig}
\usepackage{boxedminipage}
\usepackage{multirow}

\usepackage{fullpage}
\usepackage{amsmath}
\usepackage{amssymb}
\usepackage{graphicx}
\usepackage{mathrsfs}
\usepackage{wrapfig}
\usepackage{boxedminipage}
\usepackage{epsfig}
\usepackage{setspace}
\usepackage{float}
\usepackage{hyperref}
\usepackage{enumerate}
\usepackage{cite}
\usepackage[font=footnotesize,labelfont=rm]{caption}

\usepackage[numbers,sort&compress]{natbib}

\def\be{\begin{equation}}
\def\ee{\end{equation}}
\def\bea{\begin{eqnarray}}
\def\eea{\end{eqnarray}}

\numberwithin{equation}{section}

 \newcommand{\RN}[1]{%
   \textup{\uppercase\expandafter{\romannumeral#1}}%
 }
\begin{document}
\thispagestyle{empty}

\vskip 2cm

\begin{center}
{\Large \bf Shadow Thermodynamics of AdS Black Hole with the Nonlinear Electrodynamics Term}
\end{center}

\vskip .2cm

\vskip 1.2cm

\centerline{ \bf He-Bin Zheng$a$\footnote{zhenghb3060@163.com}, Ping-Hui Mou$^a$\footnote{mph2022@163.com}, Yun-Xian Chen$^a$\footnote{cyx17765580321@163.com} and Guo-Ping Li$^a$\footnote{Corresponding author: gpliphys@yeah.net}
}

\vskip 7mm
\begin{center}{ $^{a}$ School of Physics and Astronomy, China West Normal University, Nanchong 637000, People's Republic of China}	
\end{center}

\vskip 1.2cm
\centerline{\bf Abstract}
\noindent
In this paper, we have creatively employed the shadow radius to study the thermodynamics of a charged AdS black hole with a nonlinear electrodynamics(NLED) term. First, the connection between the shadow radius and event horizon is constructed with the aid of the geodesic analysis. It turns out that the black hole shadow radius shows a positive correlation as a function of the event horizon radius. Then in the shadow context, we found that the black hole temperature and heat capacity can be presented by the shadow radius. And further analysis shows that the shadow radius can do as well as the event horizon in revealing black hole phase transition process. In this sense, we constructed the thermal profile of the charged AdS black hole with inclusion of the NLED effect. In $P<P_c$ case, it is found that the $N$-type trend of the temperature given by the shadow radius is always coincide with that obtained by using the event horizon. So, we can concluded for the charged AdS black hole that the phase transition process can be intuitively presented as the thermal profile in the shadow context. Finally, the effects of NLED have been carefully analysed through out the paper.

\vskip 0.5cm

\newpage
\setcounter{footnote}{0}

\renewcommand{\baselinestretch}{1.5}
\section{Introduction} \label{Introduction}

Black hole is a very mysterious celestial bodies in the universe, which predicted by Einstein's general theory of relativity.
At present, more and more astronomers and physicists devoted themselves to the study of black hole physics both theoretically and
experimentally.
On the theoretical side, one has proved that black hole is a thermodynamic system, which possess the temperature, entropy, the negative heat capacity and so on. In this side, one has obtained a series of achievements that greatly promoted the development of black hole physics.
On the experimental side, the Laser Interferometer Gravitation-Wave Observatory (LIGO) detected the gravitational wave signal produced by the merger of two massive black holes, which provide an effective evidence for black holes.\textsuperscript{\cite{1}}
Later, the Event Horizon Telescope (EHT) has reported images of the supermassive black hole M87$^{*}$ and SgrA$^{*}$, which give a direct evidence of existence of black holes.\textsuperscript{\cite{2,3,4,5,6,7}} From those images, it shows that there always exists a dark central area with a bright ring outside, where the dark region and bright ring are now the so-called ``black hole shadow" and ``photon sphere" respectively.

Due to the strong gravitational field, the light rays came from the infinity will be deflected in the neighborhood of black hole. And, as long as the curvature of the ray approach to the critical curve, the photon will be asymptotically close to the bound photon orbit. Therefore, when the light rays emitted from a light source behind a black hole, the static observer at infinity can observe that the photons with a radius less than the bound photon orbit will fall into the black hole, which then naturally result into a shadow,\textsuperscript{\cite{8}} while other photons would move to the infinity.
This bound orbit give rise to the so called photon sphere of black hole.
The size and shape of the shadow is determined by the photon sphere. In the case of Schwarzschild black holes, photon sphere is located at $r = 3M$.\textsuperscript{\cite{9}}
Black hole shadows can give us information about the fundamental properties of black holes. Synge and Luminet were the first to study the photon trajectories of Schwarzschild black holes.\textsuperscript{\cite{10}} And later, Bardeen further studied the shadow of the rotating Kerr black hole.\textsuperscript{\cite{11}} Especially in 2017, the results inferred by EHT observations are shown to be consistent with theoretical expectations, thereby further provides a obvious  evidence for the existence of black holes in universe.\textsuperscript{\cite{12}} In recent years, many researchers have been attracted to study the black hole shadows. For instance, one finds that the shadow can be used to test the Lorentz symmetry.\textsuperscript{\cite{13}} And by considering accretions surrounded black hole or not, the shadows of different types of black holes and wormholes have also been clearly studied,\textsuperscript{\cite{14,15,16,17,18,19,20,21,22,23,24,25,26,27,28,29,30,31}} where some important result are obtained.

Besides, since black hole is regarded as a thermodynamic system, many researches have devoted to study its special thermodynamic properties.
In fact, the real understanding of the thermal properties of black holes came after the establishment of the four laws of black hole thermodynamics and the discovery of Hawking radiation.
Through quantum field theory, Hawking demonstrated that particles can emitted from the event horizon of black hole due to the vacuum fluctuation,\textsuperscript{\cite{32}} which laid a solid foundation for the study of black hole thermodynamics.
Under the premise that the ``cosmic supervision hypothesis'' and ``strong energy conditions'' are true, Hawking proved the ``Hawking area theorem'', that is, ``the area of a black hole never decreases in a clockwise direction''.
On this basis, Bekenstein considers the area and surface gravity of the event horizon to be the entropy and temperature of a black hole.
Thus, the four laws of black hole thermodynamics gradually formed.
At present, there are two main focuses on black hole thermodynamics, one is to establish the basis of the traditional four laws of black hole thermodynamics and the study of thermodynamic quantities, anther is to study black hole thermodynamics in the extended phase space, i.e., regarding the negative cosmological constant as one of the thermodynamic quantities to expand the space of macroscopic states.
For instance, the author R.B. Mann $et$ $al.$ links the negative cosmological constant to the pressure in thermodynamic systems, which has attracted many people to continue to explore this vision in depth.\textsuperscript{\cite{33}}
Later, one finds that the thermodynamic system of a charged AdS black hole is very similar to the van der Waals liquid-gas system.\textsuperscript{\cite{33,34,35,36,37,38}}
Meanwhile, Wei was the first to study the dynamic phase behavior of black hole triplic points\textsuperscript{\cite{39}} in the extended phase space.
More importantly, the relation between the thermodynamic phase structure of black holes and the shadow radius have been preliminary studied in Ref.\cite{40,41}.
In a word, both black hole thermodynamics and shadow have always been widely studied in recent years.\textsuperscript{\cite{33,34,35,36,37,38,39,40,41,42,43,44,45,46,47,48,49,50,51,52,53,54,55}}

It should be noted that linking the shadow of black hole with its thermodynamics has an important significance as it can be regarded as a bridge to reflect more information about the black hole thermodynamics.
In view of this, the purpose of this paper is to construct the relationship between the shadow radius and the thermodynamic phase transition in the AdS spacetime. We will first study the relationship between the critical thermodynamic phase transition and the black hole shadow.
Through a carefully discussion, it is believed that the black hole shadow can indeed be regarded as a powerful tool during revealing the thermodynamic phase transition process of black hole. In Ref.\cite{56}, we preliminarily find that the dependence of the black hole shadow and thermodynamics may be structured in the regular spacetime.

On the other hand, the NLED effect in curved spacetime is very important since photons propagate along geodesic lines that are no longer zero in actual Minkowski spacetime, but in corrected effective geometry\textsuperscript{\cite{57}} in this case. And, many studies have been devoted to this area accompany with much fruits.\textsuperscript{\cite{57,58,59}} In 2000, the geometrical aspects of light propagation in nonlinear electrodynamics has been discussed by Novello $et$ $al$.\textsuperscript{\cite{57}} The Novello’s method has been apply to derive the effective geometry induced by Euler-Heisenberg NLED effects in regular black
hole spacetime.\textsuperscript{\cite{58}} Also, the NLED effects has been taken into account during the study of the optical appearance of black holes.\textsuperscript{\cite{59}} However, it should be noted that the relationship between the shadow radius and thermodynamic for a NLED black hole remains unknown.
In view of this, we take the AdS black hole with the NLED term as an example to study thermodynamic phase transition with the shadow radius in this paper.

The rest of this paper are organized as follows. In Sec.2, the shadow radius of the AdS black hole with the NLED term is presented, and the critical condition of the black hole phase transition is obtained. In Sec.3, the thermodynamic phase transition of black hole is analyzed by using the shadow radius. In Sec.4, the thermal profiles of the black holes in several representative sets of coupling constants are established. Sec.5 is the conclusions and discusss. In this paper, we use the units $G_{\rm N}=\hbar=\kappa_{\rm B}=c=1$.

\section{Shadow radius and critical state} \label{BW}

Considering the action of NLED theory which minimal coupled to gravity is\textsuperscript{\cite{60}}
\begin{equation}
	\label{eq: 2.1}
	S=\frac{1}{16\pi}\int\sqrt{-g}\left[ R+K(\psi)\right] d^{4} x,
\end{equation}
with
\begin{equation}
	\label{eq: 2.2}
	\psi=F_{\rm \mu\nu}F^{\mu\nu},
\end{equation}
\begin{equation}
	\label{eq: 2.3}
	F_{\rm \mu\nu}=\bigtriangledown_{\rm \mu}A_{\rm \nu}-\bigtriangledown_{\rm \nu}A_{\rm \mu}.
\end{equation}
In Eq.{\ref{eq: 2.1}}, $R$ is the Ricci scalar, $g$ is the metric tensor determinant and $A_{\rm \mu}$ is the Maxwell field, $K(\psi)$ is the function of $\psi$. While, $\mathcal{L}(\mathcal{F})$ is the NLED Lagrangian, which depending on electromagnetic invariants, where $\mathcal{F}=\frac{1}{4}F_{\rm \mu\nu}F^{\mu\nu}$, $F^{\mu\nu}$ denots the eletromagnetic field strength tensor. The Lagrangian density of the black hole is
\begin{equation}
	\label{eq: 2.4}
	\mathcal{L}(\mathcal{F})=8\sqrt{2}\zeta\sqrt{-4\mathcal{F}}-16\mathcal{F}-8\varLambda.
\end{equation}
To study the shadow of black hole, we need to know how photons propagate in this spacetime. Note that, the effective geometry induced by the NLED, will correct the background geometry along the null geodesics of which photons would usually propagate. In our analysis, we consider that in a non-linear electrodynamic term, photons propagate along the null geodesics in effective geometry. The propagation path of photons can be described by null geodesics in effective space-time, that is\textsuperscript{\cite{57,58,61,62}}
\begin{equation}
	\label{eq: 2.5}
	g{_{eff}^{\mu\nu}}=\mathcal{L}{_\mathcal{F}}g^{\mu\nu}-\mathcal{L}{_\mathcal{F}}{_\mathcal{F}}F{_\alpha^\mu}F^{\alpha\nu},
\end{equation}
where $\mathcal{L}_\mathcal{F}\equiv\frac{\partial\mathcal{L}}{\partial\mathcal{F}}$. The effective geometry of the metric can be rewritten as
\begin{equation}
	\label{eq: 2.6}
	ds_{eff}^{2}=H(r)\left[ -f(r)dt^{2}+\frac{dr^{2}}{f(r)}\right] +h(r)r^{2}d\varOmega^{2},
\end{equation}
in which $H(r)=-\mathcal{L}_{\mathcal{F}}$ and $h(r)=-\mathcal{L}{_\mathcal{F}}{_\mathcal{F}}\frac{Q^{2}}{r^{4}}$ for metric (\ref{eq: 2.23}).
Therefore, we have\footnote{Eq.(2.7) is obtained for matric (\ref{eq: 2.23}).}
\begin{equation}
	\label{eq: 2.7}
	H(r)=1+\frac{r^{2}\zeta}{Q},~~~~~	h(r)=1+\frac{2r^{2}\zeta}{Q}.
\end{equation}
Note that the function $H(r)$ and $h(r)$ must be positive, because we want the photon not to change its own characteristics during motion.\textsuperscript{\cite{59}} Based on (\ref{eq: 2.6}), the Lagrangian of density $\mathcal{L}$ is,
\begin{equation}
	\label{eq: 2.8}
	\mathcal L=\frac{1}{2}\left[ H(r)(f(r)\dot{t}^{2}-f(r)^{-1}\dot{r}^{2})-h(r)r^{2}(\dot{\theta}^{2}+sin^{2}\theta\dot{\phi}^{2})\right] ,
\end{equation}
where $\dot{x}\equiv\frac{dx}{d\tau}$, the $\tau$ is the affine parameter. Since we only consider the photons that move on the equatorial plane ($\theta=\pi/2$, $\dot{\theta}$=0 and $\ddot{\theta}$=0). And meanwhile, the metric function does not contain time $t$ and azimuthal angle $\phi$. So, the associated equation of conserved constants can be obtained, which are
\begin{eqnarray}
	\label{eq: 2.9}
	p_{t}=\frac{\partial\mathcal{L}}{\partial\dot{t}}=H(r)f(r)\dot{t}=E,
\end{eqnarray}
\begin{equation}
	\label{eq: 2.10}
	p_{\phi}=\frac{\partial\mathcal{L}}{\partial\dot{\phi}}=h(r)r^{2}\dot{\phi}sin^{2}\theta=L,
\end{equation}
where, $E$ and $L$ represent the energy and angular momentum of photon. The four-velocity of the time, the azimuthal angle, and the radial components can be obtained,
\begin{equation}
	\label{eq: 2.11}
	\frac{dt}{d\tau}=\frac{1}{bH(r)f(r)},
\end{equation}
\begin{equation}
	\label{eq: 2.12}
	\frac{dr}{d\tau}=\sqrt{\frac{1}{b^{2}H(r)^{2}}-\frac{f(r)}{r^{2}H(r)h(r)}},
\end{equation}
\begin{equation}
	\label{eq: 2.13}
	\frac{d\phi}{d\tau}=\pm\frac{1}{r^{2}h(r)},
\end{equation}
in which the symbol ``$\pm$'' indicates the clockwise $(+)$ and counterclockwise $(-)$ direction of the motion of photon. The $b$ is the impact parameter, satisfying $b=\frac{|L|}{E}$. The equations of motion for the null geodesic is
\begin{equation}
	\label{eq: 2.14}
	H(r)f(r)\dot{t}^{2}-H(r)f(r)^{-1}(\frac{dr}{d\phi})^{2}\dot{\phi}^{2}-r^{2}h(r)\dot{\phi}^{2}=0.
\end{equation}
The above three formulas report a complete description of the photon dynamics around the black hole, where the effective potential can be written as
\begin{equation}
	\label{eq: 2.15}
	(\frac{dr}{d\tau})^{2}+V_{eff}(r)=0,
\end{equation}
with the effective potential is
\begin{equation}
	\label{eq: 2.16}
	V_{eff}(r)=\frac{1}{r^{2}H(r)h(r)}f(r)-\frac{E^{2}}{L^{2}H(r)^{2}}.
\end{equation}
It is obvious that $V_{eff}(r)$ determines the position of an unstable orbit. This position should satisfy the conditions
\begin{equation}
	\label{eq: 2.17}
	\mathcal{V}_{\rm eff}(r)=0,~~~~\mathcal{V}'_{\rm eff}(r)=0,~~~~\mathcal{V}''_{\rm eff}(r)>0.
\end{equation}
After substituting Eq.{\ref{eq: 2.16}} into Eq.{\ref{eq: 2.17}}, then we have $\frac{L}{E}=\frac{r_{p}}{\sqrt{f(r_{p})}}$, where $r_{p}$ is the radius of the photon ring. Based on Eq.{\ref{eq: 2.12}} into Eq.{\ref{eq: 2.13}}, one can obtain
\begin{equation}
	\label{eq: 2.18}
	\frac{dr}{d\phi}=(\frac{dr}{d\tau})/(\frac{d\phi}{d\tau})=\pm r \sqrt{\frac{r^{2}E^{2}h(r)^{2}}{L^{2}H(r)^{2}}-\frac{h(r)}{H(r)}f(r)}.
\end{equation}
The turning point of the photon orbit being interpreted mathematically by the constraint $\frac{dr}{d\phi}\lvert_{r=\chi}=0$, so the Eq.{\ref{eq: 2.18}} can be rewritten as
\begin{equation}
	\label{eq: 2.19}
		\frac{dr}{d\phi}=\pm r \sqrt{\frac{r^{2}f(\chi)h(r)^{2}}{\chi^{2}H(r)^{2}}-\frac{h(r)}{H(r)}f(r)}.
\end{equation}
A light ray sends from a static observer placed at $r_{\rm O}$ and transmits into the past with an angle $\beta$,\textsuperscript{\cite{41,63,64}} which is
\begin{equation}
	\label{eq: 2.20}
	cot\beta=\frac{\sqrt{g_{rr}}}{\sqrt{g_{\phi\phi}}}\lvert_{ r=r_{O}}=\pm \sqrt{\frac{r{_{O}^{2}}h(r_{O})f(\chi)}{\chi^{2}H(r_{O})f(r_{O})}-1}.
\end{equation}
With the aid of Eq.{\ref{eq: 2.19}}, we have
\begin{equation}
	\label{eq: 2.21}
	sin^{2}\beta=\frac{\chi^{2}H(r_{O})f(r_{O})}{r{_{O}^{2}}h(r_{O})f(\chi)}.
\end{equation}
At the position $r_{O}$, then the shadow radius of the black hole observed by a static observer is given by
\begin{equation}
	\label{eq: 2.22}
	r_{ss}=r_{O}sin\beta=\chi\sqrt{\frac{H(r_{O})f(r_{O})}{h(r_{O})f(\chi)}}\lvert_{\chi\rightarrow r_{p}}.
\end{equation}
In the next, we will further discuss the relationship between the shadow radius and the horizon radius in a exact black hole solution. Recently, Yu and Gao have obtained the exact black hole solution to Einstein gravity with non-linear electrodynamic field, which is very different from the RN-AdS solution in Einstein-Maxwell gravity with a cosmological constant. The metric can be written as\textsuperscript{\cite{60}}
\begin{equation}
	\label{eq: 2.23}
	f(r)=1-\frac{2M}{r}+\frac{Q^{2}}{r^{2}}-\frac{\varLambda r^{2}}{3}-\frac{r^{2}\zeta^{2}}{3}+2Q\zeta,
\end{equation}
where, $\zeta$, $\varLambda$, $M$ and $Q$ are the coupling constant, cosmological constant, the mass and charge of black hole, respectively. The horizon radius is the largest root of $f(r_{h})=0$. In this case, the black hole mass can be written as a function of $r_{h}$, which is
\begin{equation}
	\label{2.24}
	M=\frac{3Q^{2}+3r{_h^2}+8P\pi r{_h^4}+6Qr{_h^4}\zeta-r{_h^4}\zeta^{2}}{6r{_h}}.
\end{equation}
And, the Hawking temperature is
\begin{equation}
	\label{eq: 2.25}
	T=\frac{f'(r)}{4\pi}\lvert_{r=r_{h}}=\frac{-Q^{2}+r{_h^2}+2Qr{_h^2}\zeta+r{_h^4}(8P\pi-\zeta^{2})}{4\pi r{_h^3}}.
\end{equation}
The state equation can be written as
\begin{equation}
	\label{eq: 2.26}
	P=\frac{Q^{2}}{8\pi r{_h^4}}-\frac{1}{8\pi r{_h^2}}+\frac{T}{2r{_h}}-\frac{Q\zeta}{4\pi r{_h^2}}+\frac{\zeta^{2}}{8\pi}.
\end{equation}
By using the metric function and the effective potential, one can obtain the photon circular orbit radius in this spacetime. It is
\begin{equation}
	\begin{split}
		\label{eq: 2.27}
		r_{\rm p} = \frac{432\pi P-54\zeta^{2}}{3\times2^{\frac{2}{3}} (8\pi P-\zeta^{2}) W}
		-\frac{W}{6\times2^\frac{1}{3}(8\pi P-\zeta^{2})},
	\end{split}		
\end{equation}
\begin{equation}
	\label{eq: 2.28}
	W\equiv\sqrt[3]{-2268\zeta^{4}M+Z-145152\pi^{2}M P^{2}+36288\pi \zeta^{2}M P},
\end{equation}
\begin{equation}
	\label{eq: 2.29}
	Z\equiv\sqrt{(-2268\zeta^{4}M-145152\pi^{2}M P^{2}+36288\pi \zeta^{2}M P)^{2}+4(432\pi P-54\zeta^{2})^{3}}.
\end{equation}
Combining Eq.{\ref{eq: 2.22}}, the shadow radius of the RN-ADS black hole with the effects of NLED can be written as
\begin{equation}
	\label{eq: 2.30}
	r_{\rm s} = r_{\rm p} \sqrt{\frac{H(r_{\rm O})f(r_{\rm O})}{h(r_{\rm O})f(r_{\rm p})}}.
\end{equation}
Combining Eq.{\ref{eq: 2.23}}, Eq.{\ref{eq: 2.27} and, Eq.{\ref{eq: 2.30}} the specific expression of the shadow radius can be obtain as Eq.{\ref{eq: 2.31}}, which is
\begin{equation}
	\label{eq: 2.31}
	r_{\rm s} = \frac{\sqrt{3}CC\times\sqrt{\frac{f(r_{\rm O})}{AA+BB}}}{2\sqrt[3]{DD}(\zeta^{2}-3)},
\end{equation}
where $AA\equiv\frac{2\pi CC^{2} P}{\sqrt[3]{DD^{2}(\zeta^{2}-3)^{2}}}-\frac{CC^{2}\zeta^{2}}{4\sqrt[3]{DD^{2}}(\zeta^{2}-3)^{2}}+\frac{12\sqrt[3]{DD^{2}}(\zeta^{2}-3)^{2}Q^{2}}{CC^{2}}$,\\ $BB\equiv\frac{2\sqrt[3]{DD}(\zeta^{2}-3)(r{_h^4}(\zeta^{2}-8\pi P)-3 Q^{2}-6\zeta Q r{_h^2}-3 r{_h^2})}{CC r{_h}}+6\zeta Q +3$, $CC\equiv-18+6\zeta^{2}+\sqrt[3]{DD^{2}}$,\\
$DD\equiv\frac{1}{r{_h}}(189(-Q^{2}-r{_h^2}-r{_h^4})-378 Q r{_h^2}\zeta+126 Q^{2}\zeta^{2}+126 r{_h^2}\zeta^{2}+189 r{_h^4}\zeta^{2}+252 Q r{_h^2}\zeta^{3}-21 Q^{2}\zeta^{4}-21 r{_h^2}\zeta^{4}-63 r{_h^4}\zeta^{4}-42 Q r{_h^2}\zeta^{5}+7 r{_h^4}\zeta^{6}+\frac{r{_h}}{27}\sqrt{\frac{r{_h^2}(162-54\zeta^{2})^{3}+35721(-3+\zeta^{2})^{4}(3Q^{2}+3r{_h^2}+6Qr{_h^2}\zeta-r{_h^4}(-3+\zeta^{2}))^{2}}{r{_h^2}}})$.\\
Using the state Eq.{\ref{eq: 2.26}} and the critical condition
\begin{equation}
	\label{eq: 2.32}
	({\partial P}/{\partial r_{\rm h}})=({{\partial}^{2}P}/{\partial r_{\rm h}^{2}})=0,
\end{equation}
one can get
\begin{eqnarray}
	\label{eq: 2.33}
	&&P_{\rm c}=\frac{1+4Q\zeta+16Q^{2}\zeta^{2}}{96\pi Q^{2}},\\
	\label{eq: 2.34}
	&&r_{\rm c}=\frac{\sqrt{6}Q}{\sqrt{1+2\zeta}},\\
	\label{eq: 2.35}
	&&T_{\rm c}=\frac{(1+2Q\zeta)^{3/2}}{3\sqrt{6}\pi Q}.
\end{eqnarray}

\vskip0.3cm
Due to the fact that the static observer always locates at spatial infinity, thereby we have the constraint $f(r_{\rm O})=1$ for $r_{\rm O}=100$.\textsuperscript{\cite{40}} Therefore, we have fixed the location of the observer at $r_{\rm O}=100$ in Fig.1.
The three graphs in Fig.1 describe the shadow radius of black hole as a function of the event horizon radius for different values of the coupling constant $\zeta$.
It can be seen that when the event horizon radius $r_h$ increased the shadow radius $r_s$ will increase. And, the trend of $r_s$ will be gradually flat with increase of $r_h$.
This means there exists a positive correlation between $r_h$ and $r_s$, so that one can employ the shadow radius to present the Hawking temperature of black hole.
In addition, the coupling constant decreased the event horizon radius, and meanwhile decreased the shadow radius.
When the coupling constant $\zeta \rightarrow 0$, this black hole will reduced into the RN-AdS black hole.
As $\zeta\rightarrow0$, the expression Eq.{\ref{eq: 2.31}} can be used to describe the relationship between the shadow radius and the event
horizon radius of the RN-AdS black hole.
For the RN-AdS black hole, there also exhibits the positive correlation between $r_h$ and $r_s$. So, it is believed that the shadow radius can do as well as the event horizon radius for describing the black hole temperature.
\begin{center}
	\includegraphics[width=4cm,height=4cm]{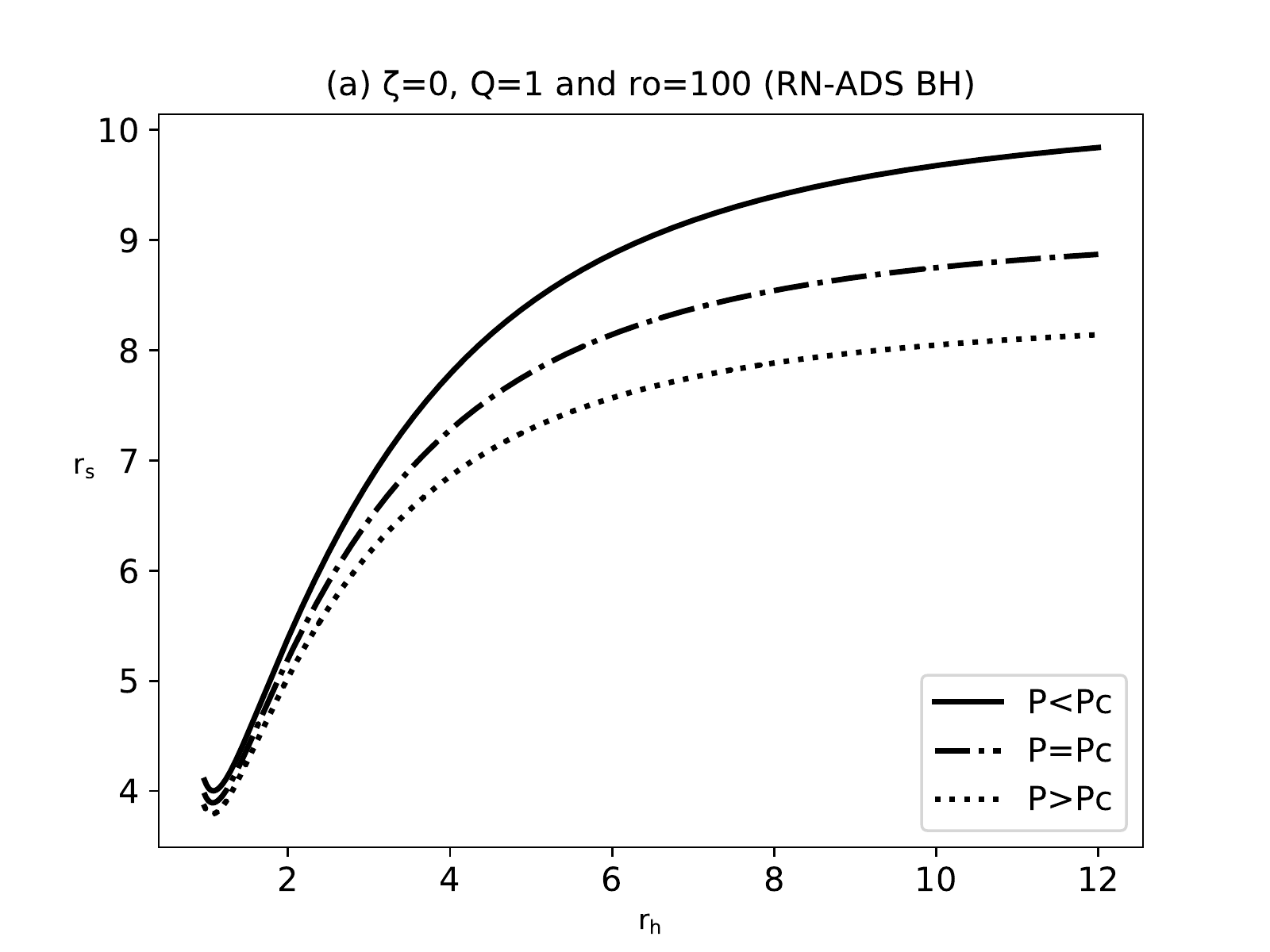}
	\includegraphics[width=4cm,height=4cm]{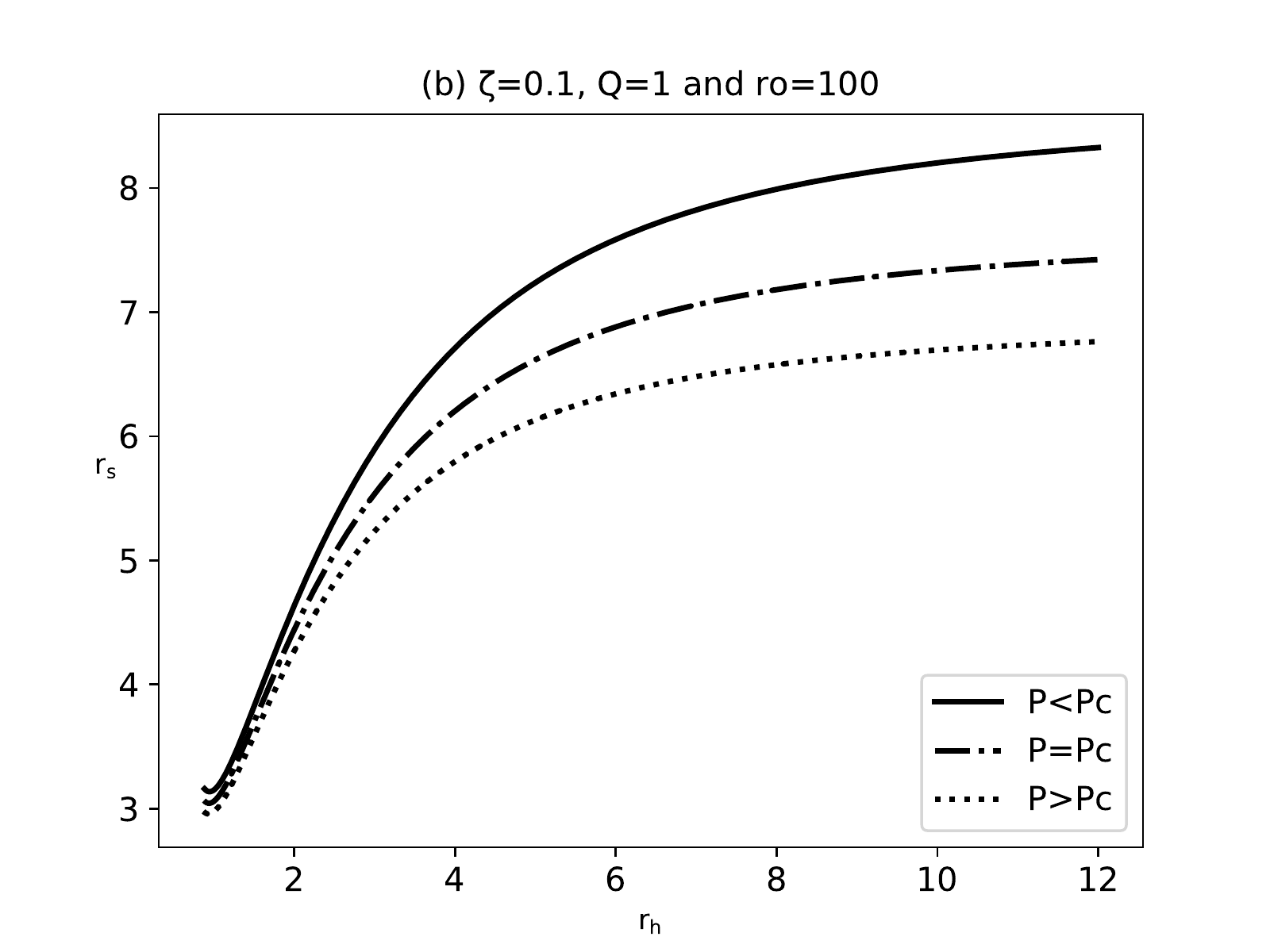}
	\includegraphics[width=4cm,height=4cm]{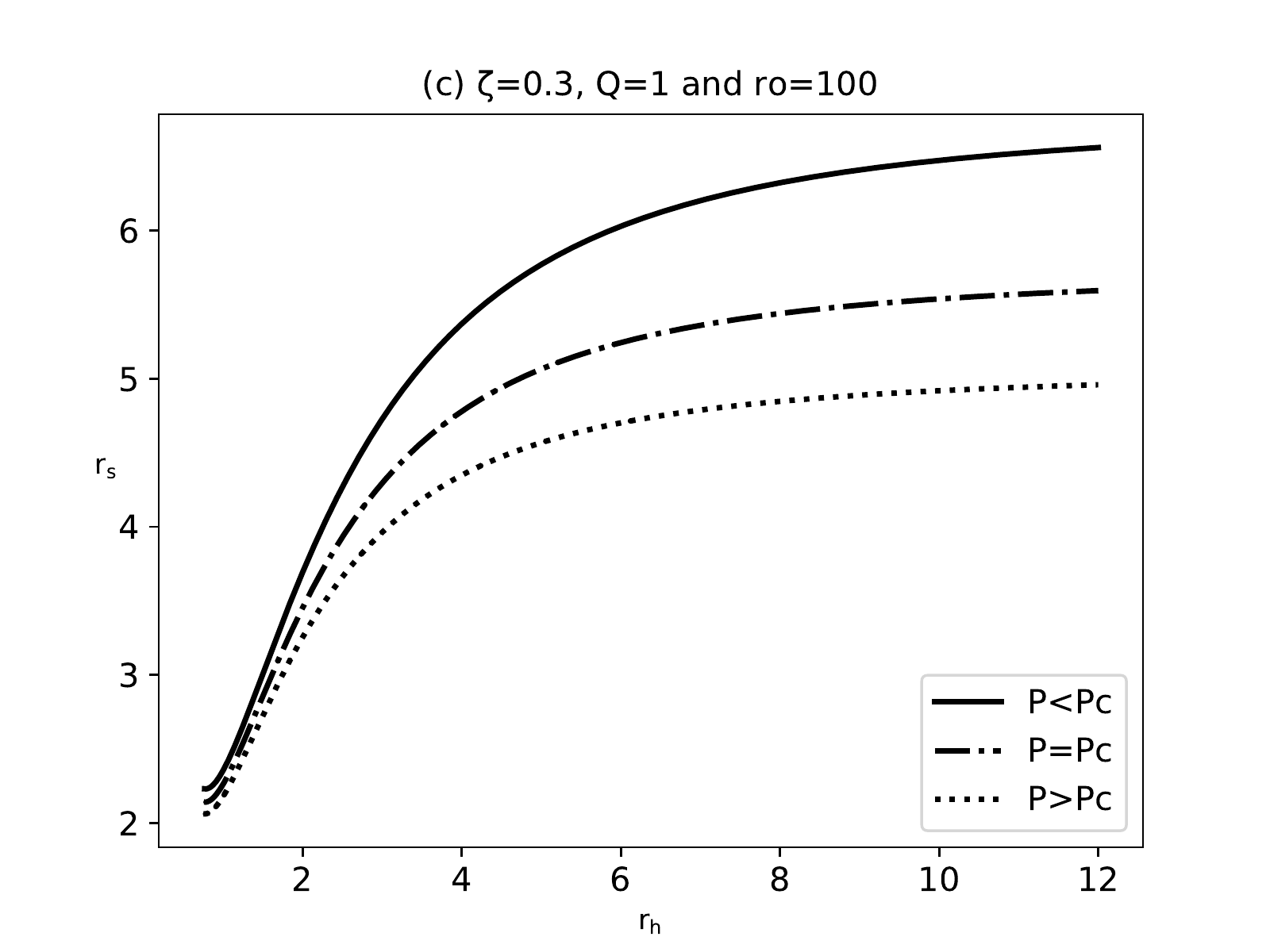}
	\parbox[c]{15.0cm}{\footnotesize{\bf Fig~1.}
		The variation of shadow radius $r_{\rm s}$ in terms of the event horizon radius $r_{\rm h}$ for the charged AdS black hole with non-linear electrodynamic field. {\em Panel (a)}-- coupling constant $\zeta=0$ and $r_{\rm O}=100$ (RN-AdS black hole), {\em Panel (b)}-- coupling constant $\zeta=0.1$ and $r_{\rm O}=100$, {\em Panel (c)}-- coupling constant $\zeta=0.3$ and $r_{\rm O}=100$. The solid lines, segment point lines and dotted lines correspond to $P<P_{\rm c}$, $P=P_{\rm c}$ and $P>P_{\rm c}$, respectively.}
	\label{fig1}
\end{center}

\section{Phase transition of a charged AdS black hole with non-linear electrodynamics term by using shadow analysis}\label{topology4}

In this section, we will further discuss the thermodynamic phase transition of the charged AdS black hole with the non-linear electrodynamics term.
According to the Hawking temperature Eq.{\ref{eq: 2.26}}, one can easily see that the black hole temperature acts as a function of the radius of the event horizon.
As long as the coupling constant is fixed to some certain value, the temperature of the black hole will vary with the radius of the event horizon, which is shown on Panel(a) of Fig.2.
Black hole temperature, as a continuous function of its event horizon radius, it exhibits different properties for different values of pressure.
In Fig.2, it can be seen that when $(P>P_{\rm c})$, the function image is a monotonically increasing smooth curve, the slope of which first decreases and then increases, but the its slope would never decrease to zero. In this case, the curve has no inflection point and black hole is in the supercritical phase.
When $(P=P_{\rm c})$, the function image is still monotonically increasing. However, the slope of it first decreases to zero, and then begin to increase.
This curve corresponds to the critical isobaric, which has an inflection point. In this case, black hole is thermodynamically unstable.
When $(P<P_{\rm c})$, the function image is no longer a monotonous change, but first increases, then decreases and finally increases.
When the pressure is lower than the critical pressure, there are two-phase transition branches. And, one is in the small radius region, which corresponds to the fluid phase of the van der Waals system, another is in the large radius region, which corresponds to the gas phase.

\begin{center}
	\includegraphics[width=4cm,height=4cm]{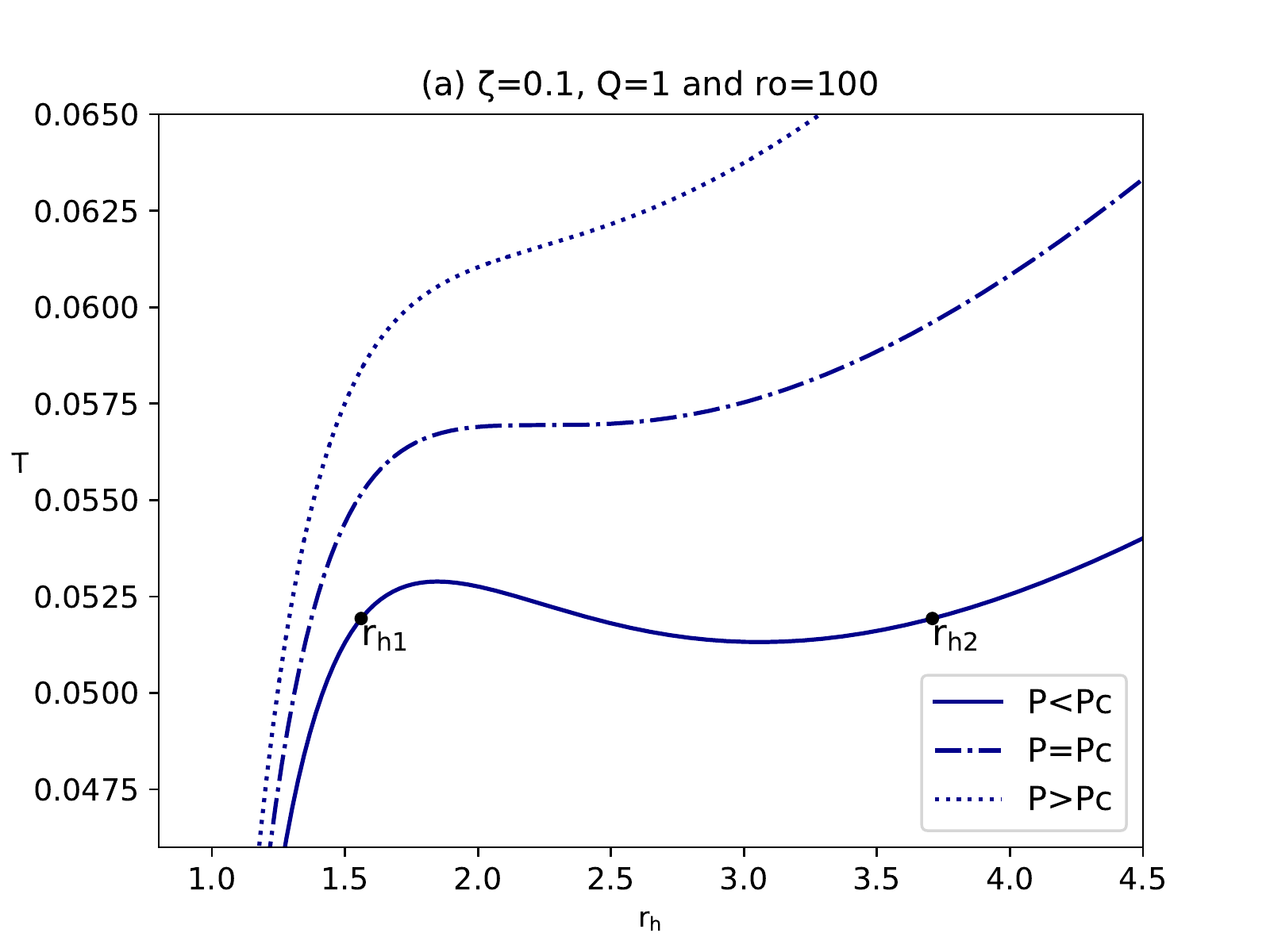}
	\includegraphics[width=4cm,height=4cm]{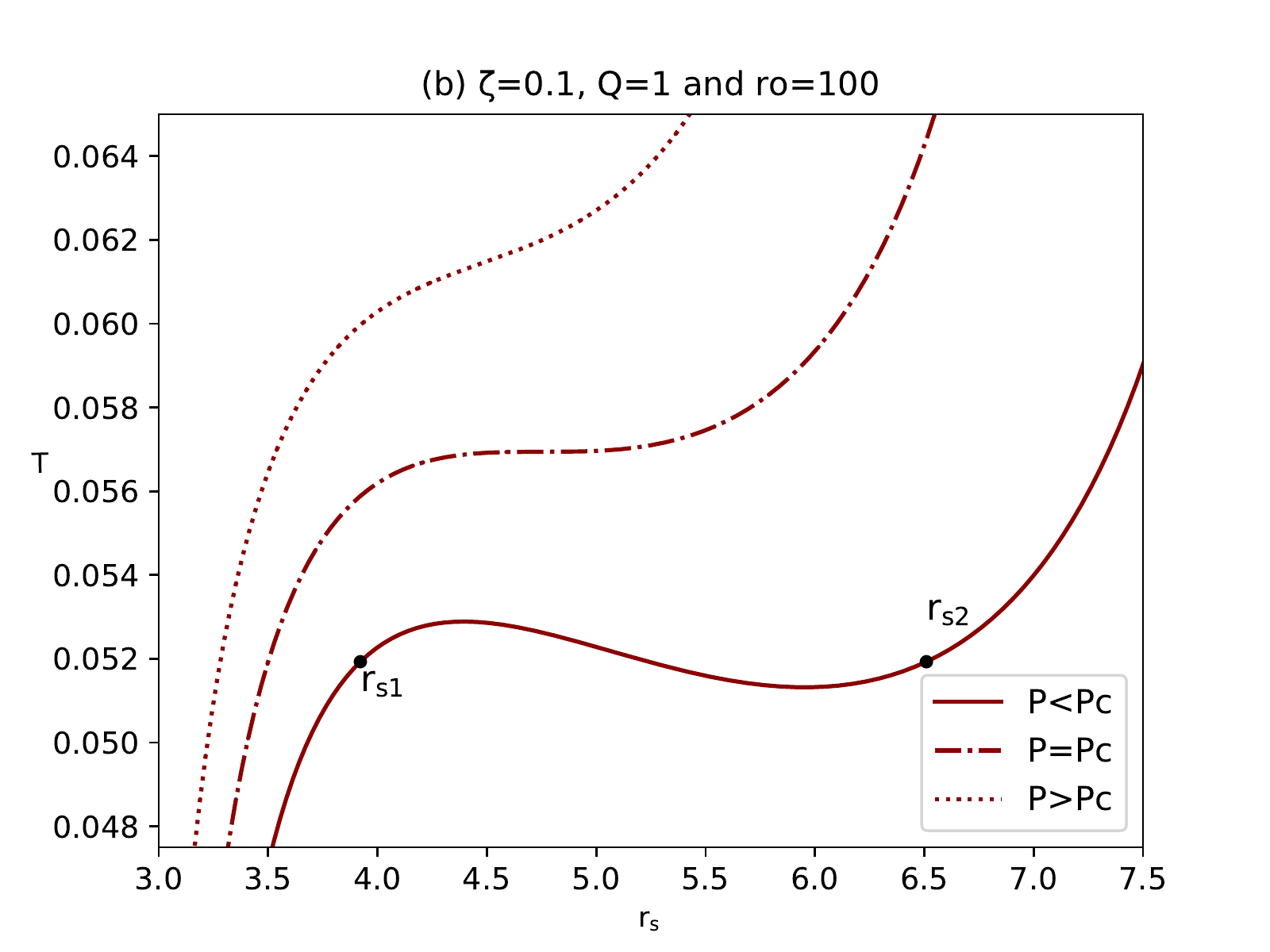}
	\includegraphics[width=4cm,height=4cm]{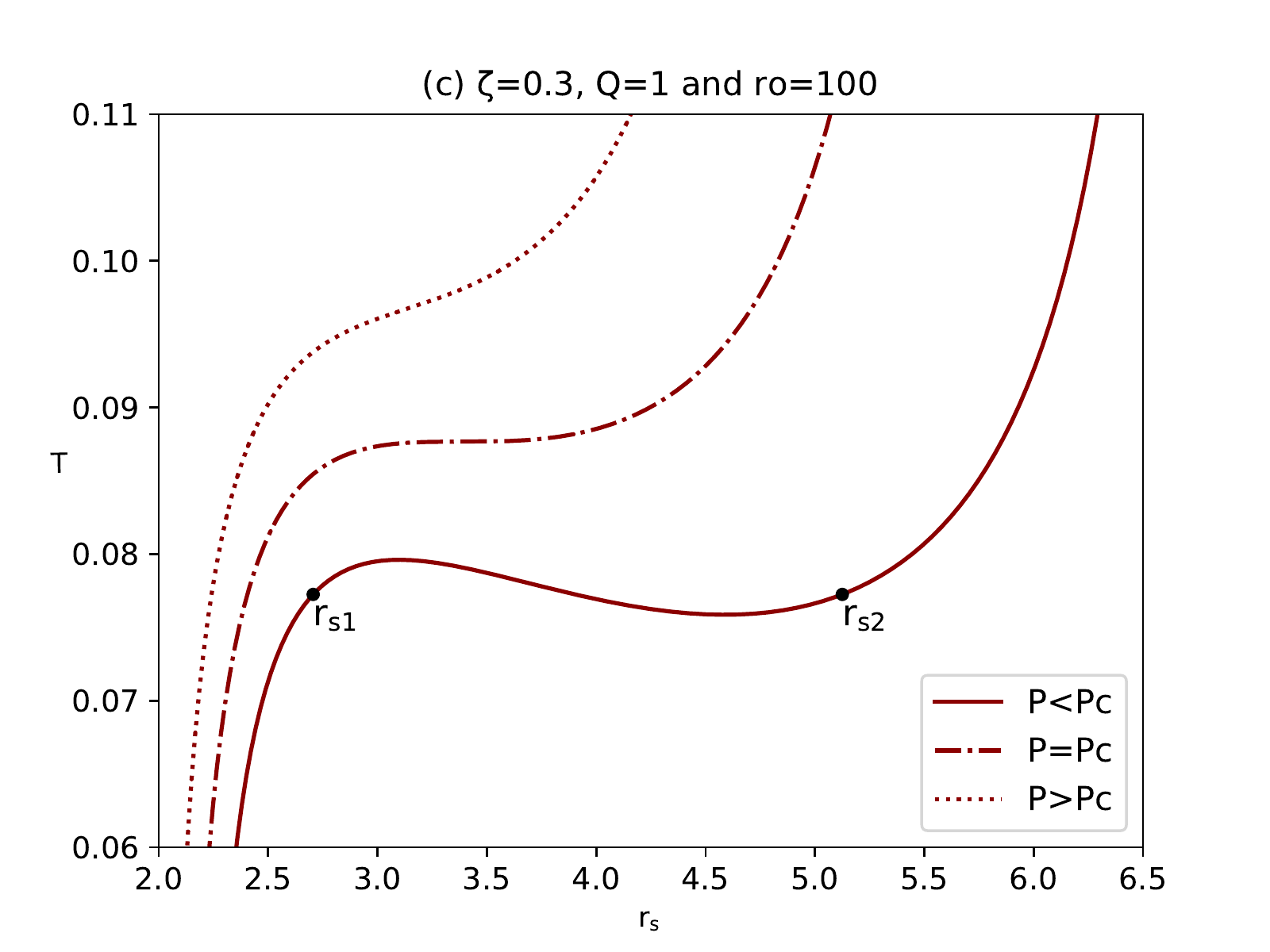}
	\parbox[c]{15.0cm}{\footnotesize{\bf Fig~2.}
		{\em Panel (a)}-- temperature as a function of $r_{\rm h}$ with $\zeta=0.1$, {\em Panel (b)}-- temperature as a function of $r_{\rm s}$ with $\zeta=0.1$, {\em Panel (c)}-- temperature as a function of $r_{\rm s}$ with $\zeta=0.3$. A static observer at $r_{\rm O}=100$.}
	\label{fig2}
\end{center}
In Panel (b) and Panel (c) of Fig.2, we present images of black hole temperature as a function of shadow radius for different values of pressure.
The Panel (b) corresponds to the coupling constant $\zeta=0.1$, while Panel (c) corresponds to the case $\zeta=0.3$.
By choosing some representative values of the coupling coefficient $\zeta$, it is true that one can employ the shadow radius instead of the event horizon to study the process of black hole phase transition.
When the pressure is less than the critical pressure, the case ($r_{\rm s}<r_{\rm s1}$) corresponds to a stable small black hole; and ($r_{\rm s}>r_{\rm s2}$), corresponds to a large stable black hole; while the case ($r_{\rm s1}<r_{\rm s}<r_{\rm s2}$) corresponds to the unstable state of black hole.
Also, we find that a larger value of $\zeta$ give rise to a stronger phase transition temperature, but a smaller value of shadow radius.
Meanwhile, Maxwell's equal area theorem can be constructed in $T-r_{\rm h}$ plane, $T_{\rm h0}(r_{\rm h2}-r_{\rm h1}) = {\int_{r_{\rm h2}}^{r_{\rm h1}} T {\rm d}r_{\rm h}}$. But when we are constructed in $T-r_{\rm s}$ plane, $T_{\rm s0}(r_{\rm s2}-r_{\rm s1}) \neq {\int_{r_{\rm s2}}^{r_{\rm s1}} T {\rm d}r_{\rm s}}$, the left side of the inequality is slightly smaller than the right side. It may be the following reason why they are not equal. As can be seen from Fig 1, the shadow radius and the radius of the event horizon are not proportional to each other, but only positively correlated.

In addition, the phase transition grade is closely related to the heat capacity of black hole. Specifically speaking, the second-order phase transition at the thermodynamic critical point of the black hole can be represented by the heat capacity mutation and specific heat diverge. By considering the relationship ($S=\frac{A}{4}=\pi r_{\rm h}^{2}$) and the temperature Eq.{\ref{eq: 2.26}}, we can obtain the heat capacity of black hole under constant pressure, i.e.,
\begin{equation}
	\label{eq: 3.3}
	C_{\rm P}=T\Bigg(\frac{dS}{dT}\Bigg)_{\rm P,\zeta}=\frac{2\pi r_{\rm h}^{2}(Q^{2}-r_{\rm h}^{2}-2Qr_{\rm h}^{2}\zeta+r_{\rm h}^{4}(-8P\pi+\zeta^{2}))}{-3Q^{2}+r_{\rm h}^{2}+2Qr_{\rm h}^{2}\zeta+r_{\rm h}^{4}(-8P\pi+\zeta^{2})}.
\end{equation}
\noindent
The Panel (a) of Fig.3 shows the heat capacity as a function of the radius of the event horizon at different critical pressures.
The heat capacities as a function of shadow radius for different critical pressures are shown in Panel (b) and Panel (c) of Fig.3.
It can be seen that the heat capacity diverges at the critical point, which implies the emergence of the second-order phase transition of black hole.
For the Panel (b) and Panel (c), the similar phenomenon of heat capacity also appears at the critical point, thereby the shadow radius can be used to reveal the black hole phase transition.
\begin{center}
	\includegraphics[width=4cm,height=4cm]{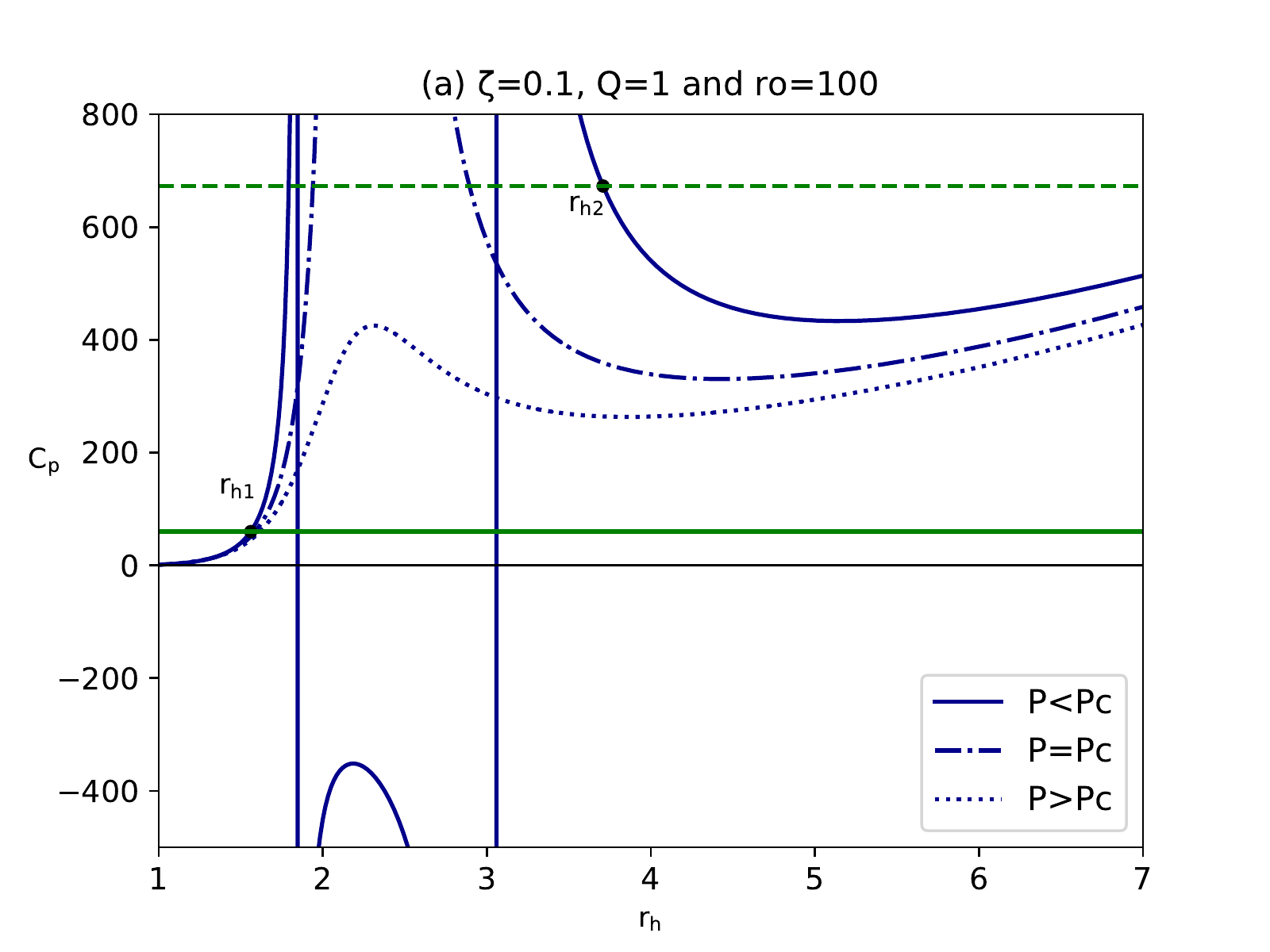}
	\includegraphics[width=4cm,height=4cm]{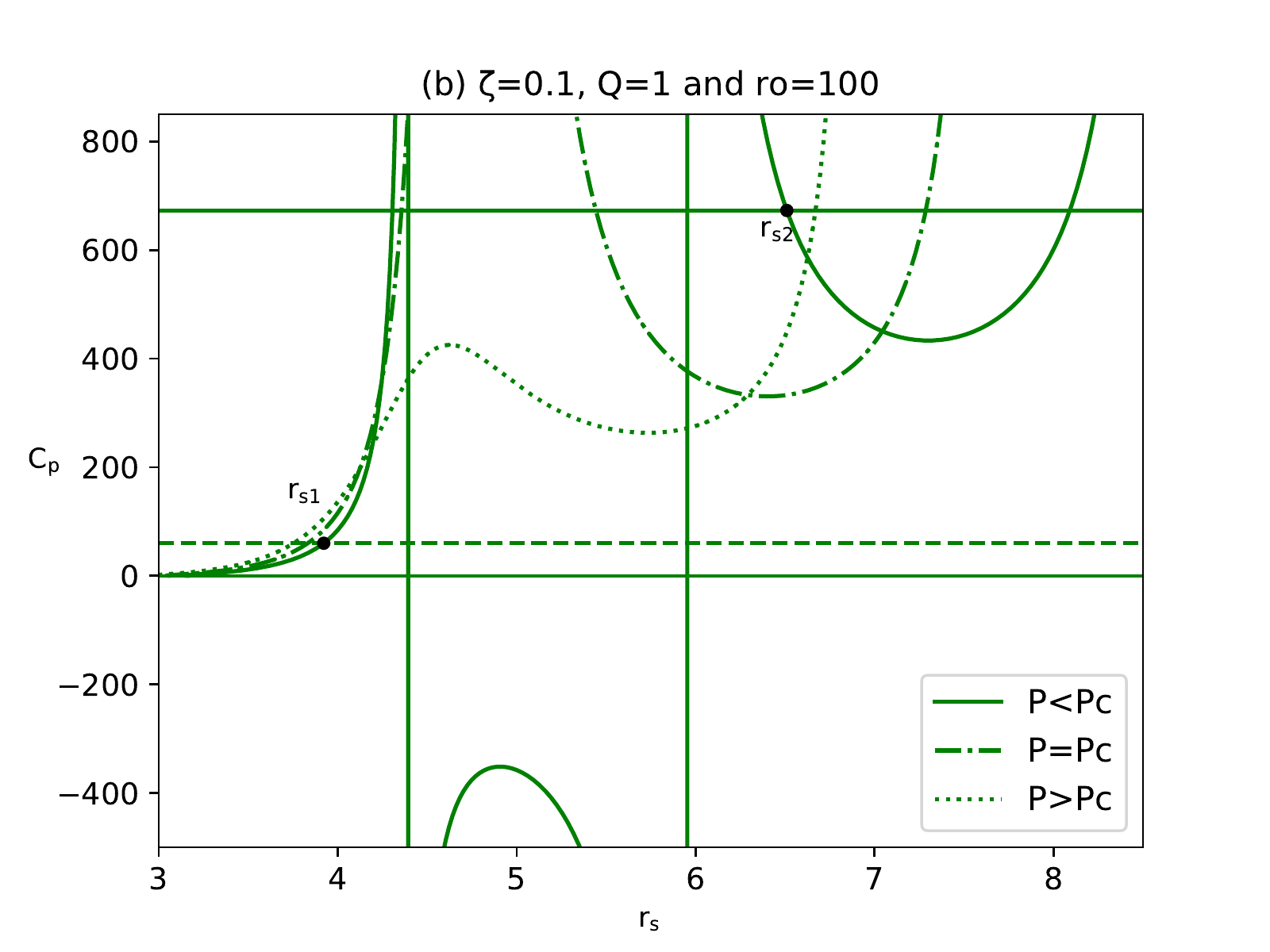}
	\includegraphics[width=4cm,height=4cm]{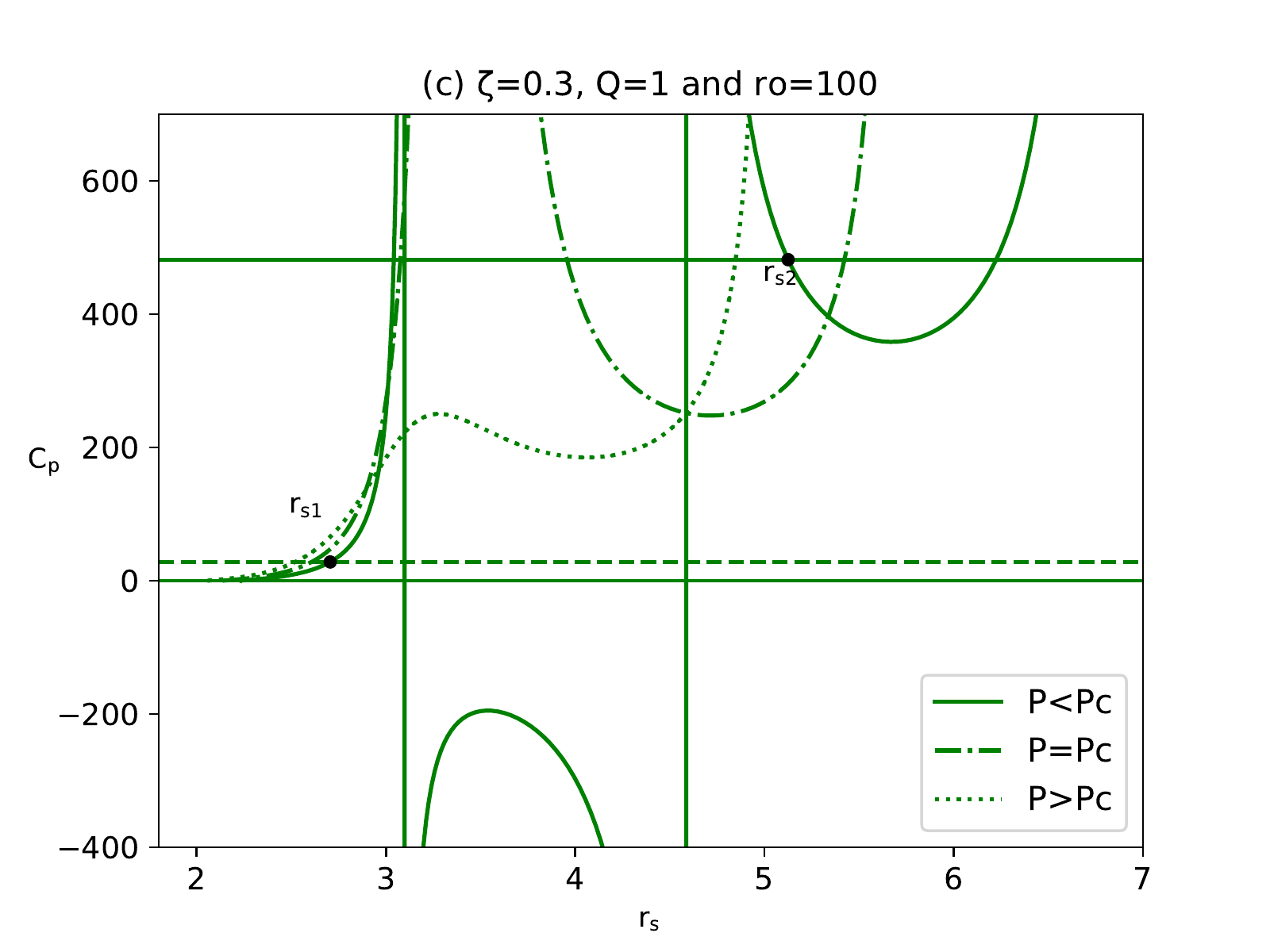}
	\parbox[c]{15.0cm}{\footnotesize{\bf Fig~3.}
		{\em Panel (a)}-- heat capacity as a function of $r_{\rm h}$ with $\zeta=0.1$, {\em Panel (b)}-- heat capacity as a function of $r_{\rm s}$ with $\zeta=0.1$. {\em Panel (c)}-- heat capacity as a function of $r_{\rm s}$ with $\zeta=0.3$. A static observer at $r_{\rm O}=100$.}
	\label{fig3}
\end{center}

\section{Thermal profile of a charged AdS black hole with non-linear electrodynamics term}\label{boundary}

In this section, we will use the thermal profile of black hole to intuitively show the relationship between the shadow and the phase structure of black hole. The shadow contour curve of celestial coordinates can be expressed by the following formula,\textsuperscript{\cite{65}} it is
\begin{eqnarray}
	\label{eq: 4.1}
	&& x = \lim\limits_{r \rightarrow \infty}\Big(-r^{2}\sin\theta_{0}\frac{{\rm d}\phi}{{\rm d}r}\Big)_{\rm \theta_{\rm 0} \rightarrow \frac{\pi}{2}},\\
	\label{eq: 4.2}
	&& y = \lim\limits_{r \rightarrow \infty}\Big(r^{2}\frac{{\rm d}\theta}{{\rm d}r}\Big)_{\rm \theta_{\rm 0} \rightarrow \frac{\pi}{2}}.
\end{eqnarray}
For a static observer, the shadow contour is shown in Fig.4.
The radius of the shadow contour is greatly affected by the pressure.
It is easy to see that as the pressure increases the shadow radius decreases.
The green point line represents $P>P_{\rm c}$, black hole in this case is at the supercritical phase.
The orange segment point line is $P=P_{\rm c}$, black hole is just at critical pressure, and the shadow in this case is more obvious compared to the supercritical case.
And the blue dotted line corresponds to $P<P_{\rm c}$, which shows the shadow profile is in the large radius region.
It is worth mentioning that as the coupling constant increased, the shadow radius decreased, which can be easily observed in Fig.4.
\begin{center}
	\includegraphics[width=5.5cm,height=5.5cm]{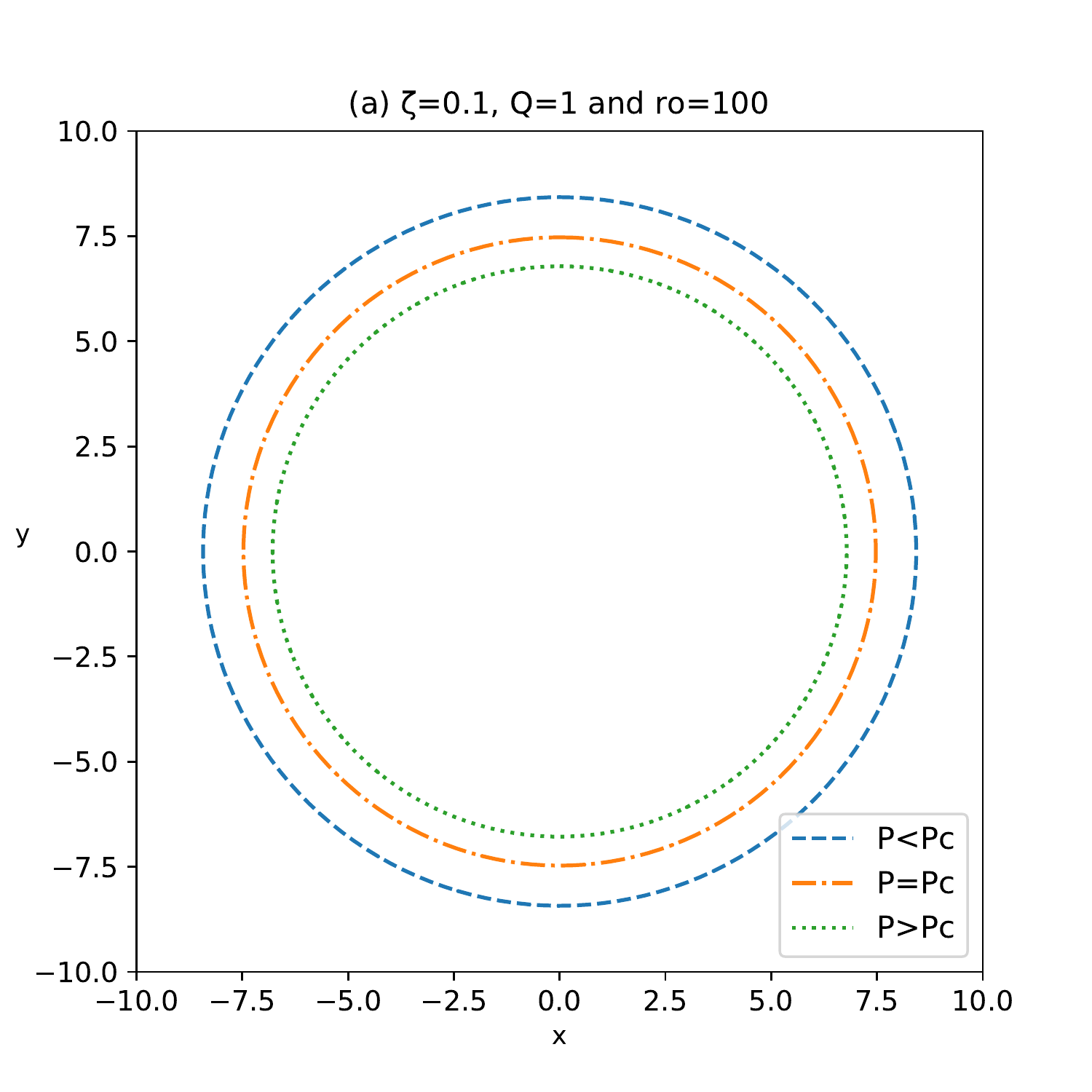}
	\includegraphics[width=5.5cm,height=5.5cm]{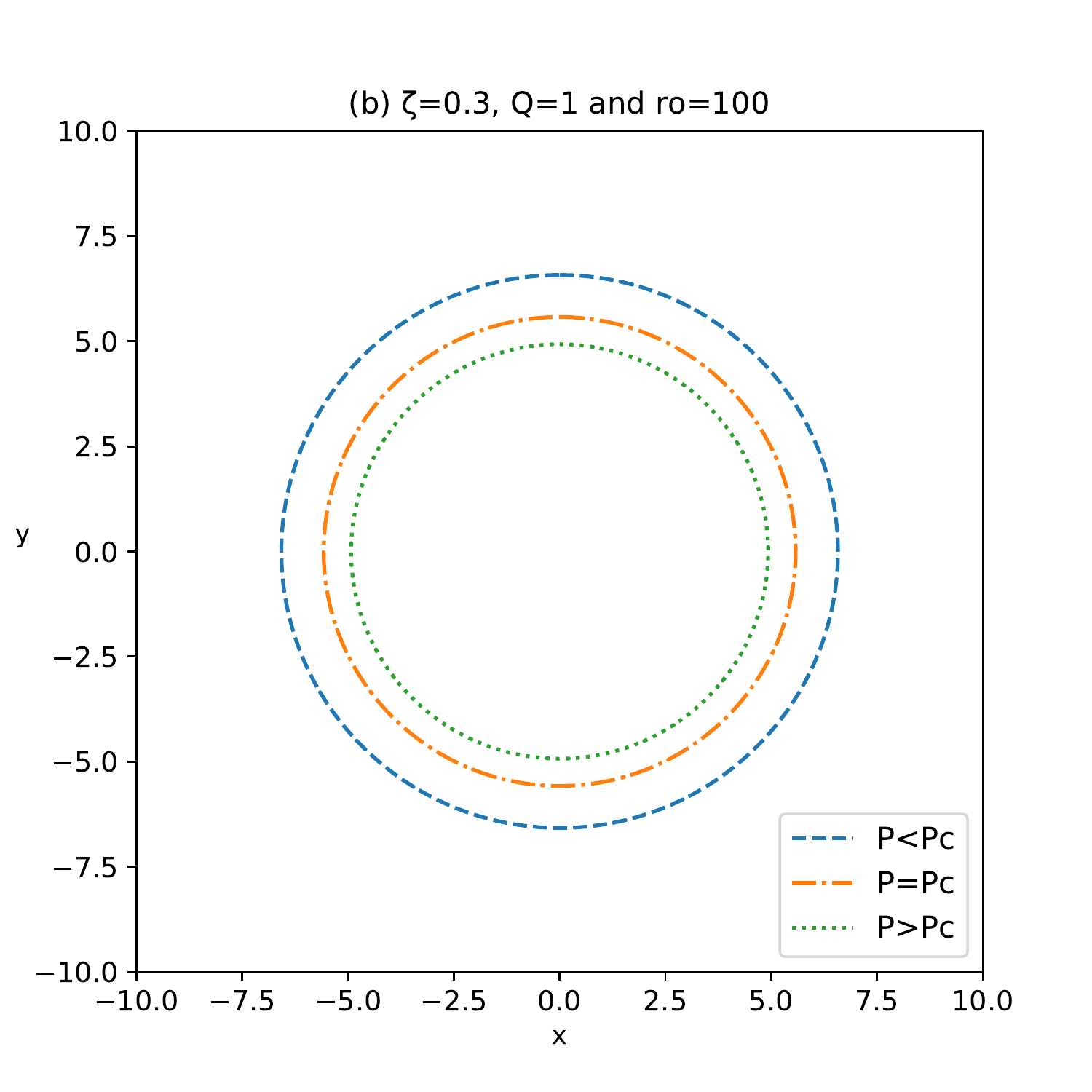}
	\parbox[c]{15.0cm}{\footnotesize{\bf Fig~4.}
		Shadow cast of  the RN-ADS BH with the effects of NLED. {\em Panel (a)}-- coupling constant $\zeta=0.1$, {\em Panel (b)}--coupling constant $\zeta=0.3$. Here, the BH mass is $M=60$ and $r_{\rm O}=100$.}
	\label{fig4}
\end{center}
Combining Fig.2 with Fig.4, we built the black hole thermal profile with the aid of the temperature-shadow radius function.
Under several representative values of coupling constant $\zeta$, Fig.5 present three different scenarios: $P>P_{\rm c}, P=P_{\rm c}, P<P_{\rm c}$.
For simplicity, the six figures in Fig.5 are marked as ($a, b, c, d, e, f$) in sequence.
Both panel (a) and (d) represent the thermal distribution of a black hole under supercritical phase $P>P_{\rm c}$.
They have a smaller shadow radius. And, the temperature of black hole gradually increases from the center of the shadow to the boundary.
This cases correspond to the dotted lines in Fig.2.
And, both panel (b) and (e) represent unstable black holes in a critical state $P=P_{\rm c}$.
In this critical region,  the black hole is thermodynamically unstable. And, the shadow radius of black hole is larger than that of the black hole in the supercritical state.
This cases correspond to the segment point lines in Fig.2.
While, Panel (c) and (f) both show black hole with the pressure is less than the critical value $P<P_{\rm c}$.
The temperature of black hole in this state has a dramatic trend from inside to outside.
The temperature of a black hole increases, then decreases, and finally increases again.
We name this change is an ``N-type'' change.
Therefore, it can be seen that the ``N-type'' change of temperature with $r_s$ in the celestial coordinate is always full consistent with the previous analysis results, i.e., Fig.2.
\begin{center}
	\includegraphics[width=4.5cm,height=4cm]{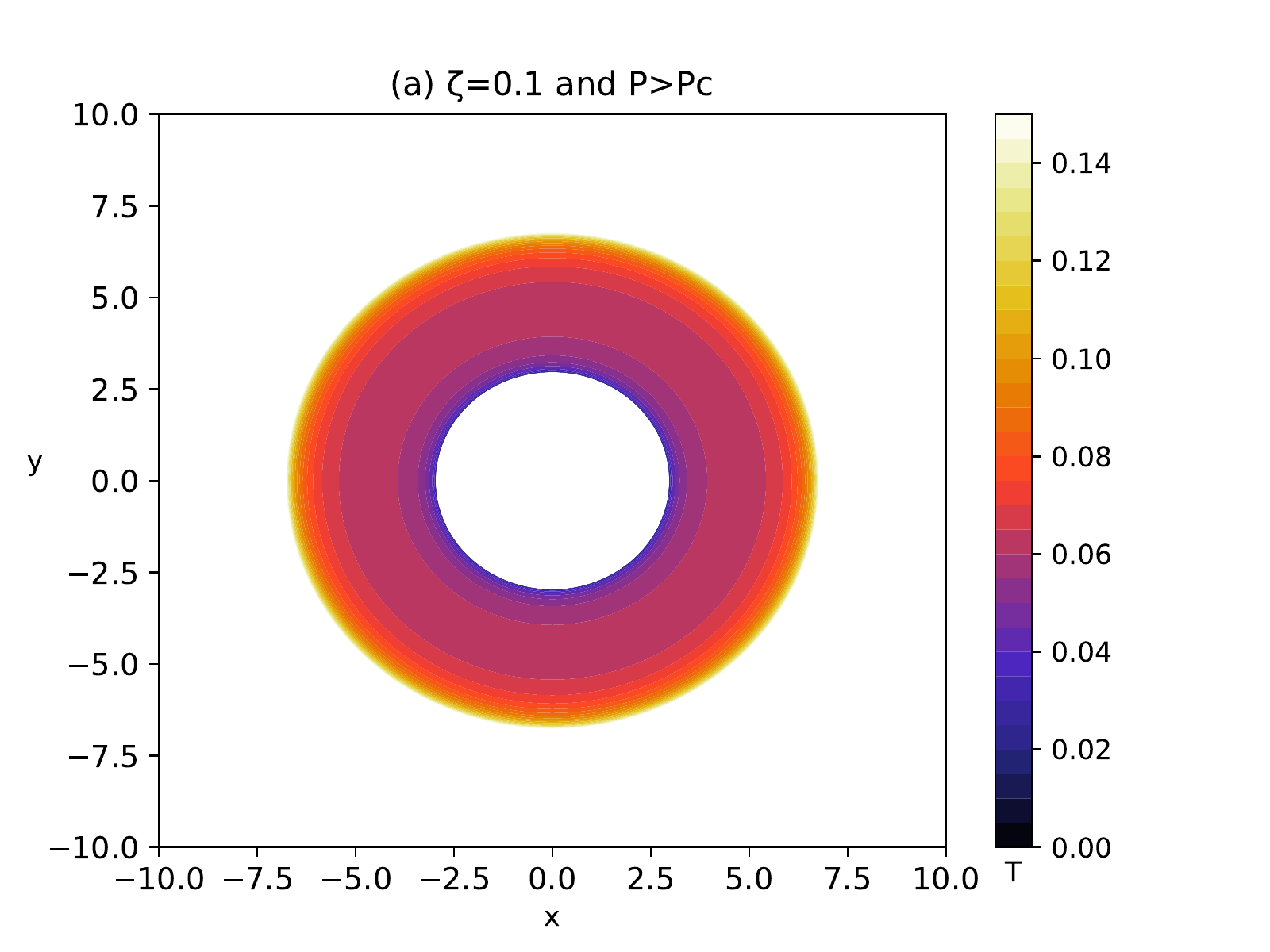}
	\includegraphics[width=4.5cm,height=4cm]{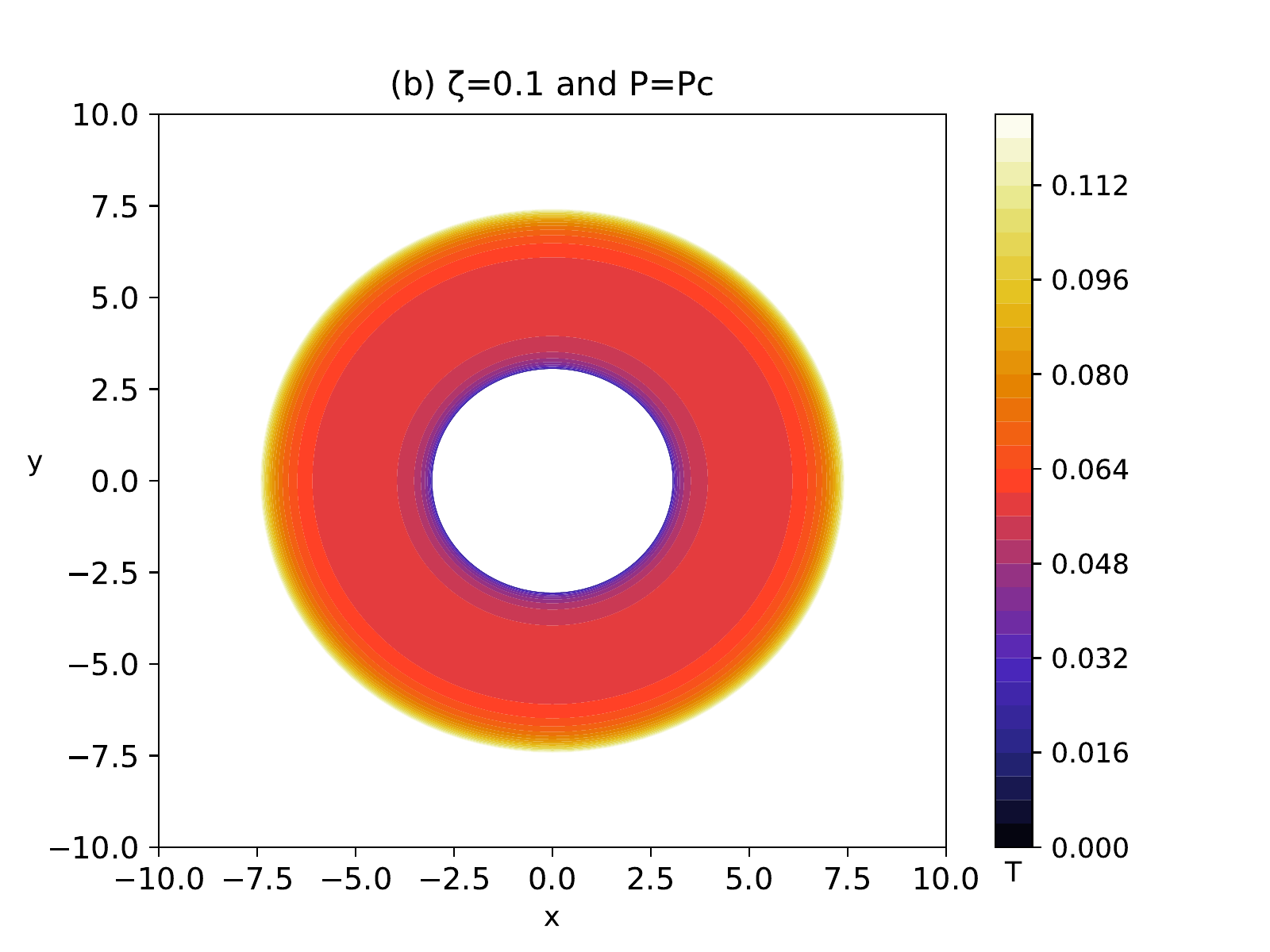}
	\includegraphics[width=4.5cm,height=4cm]{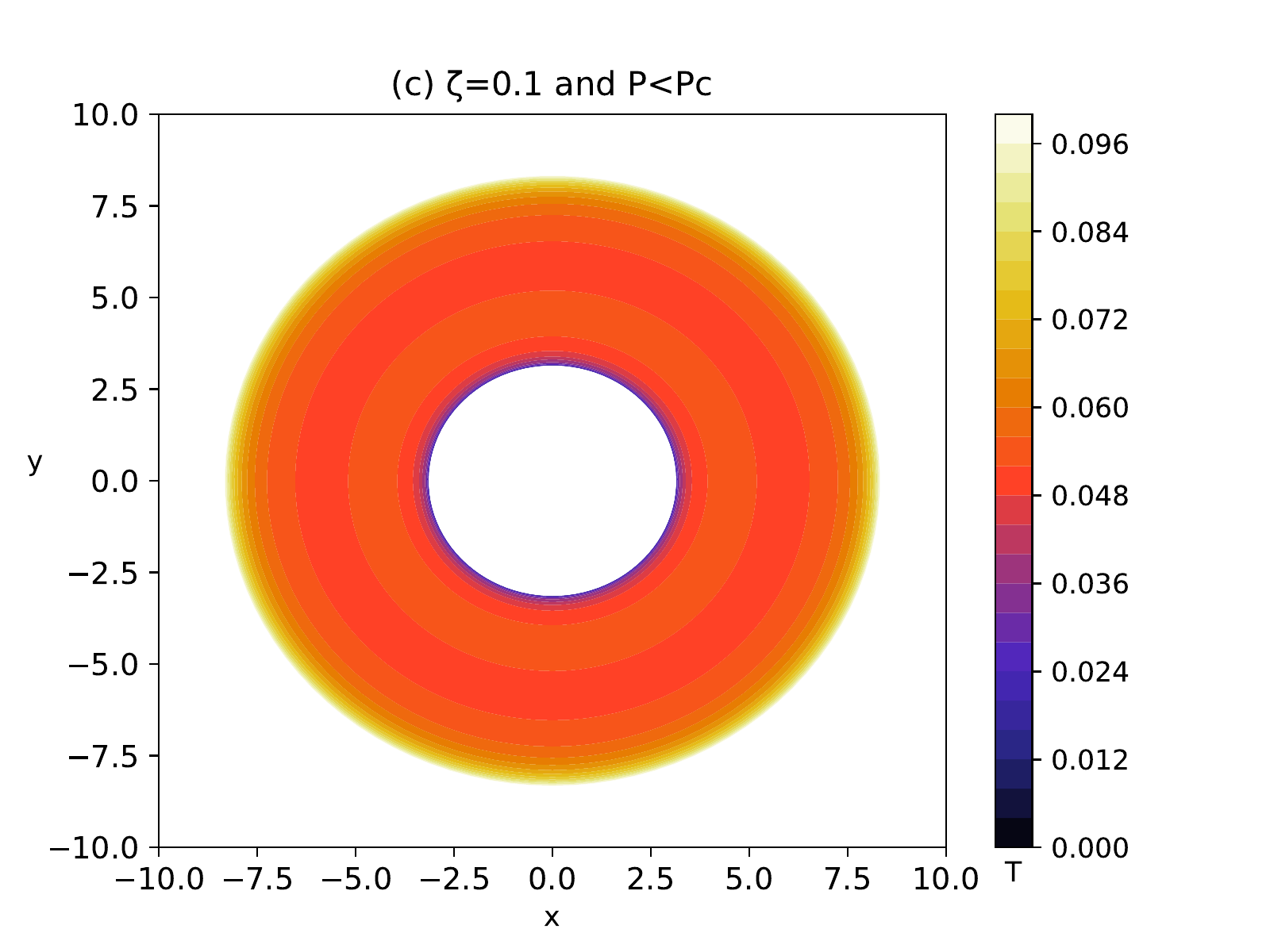}
	\includegraphics[width=4.5cm,height=4cm]{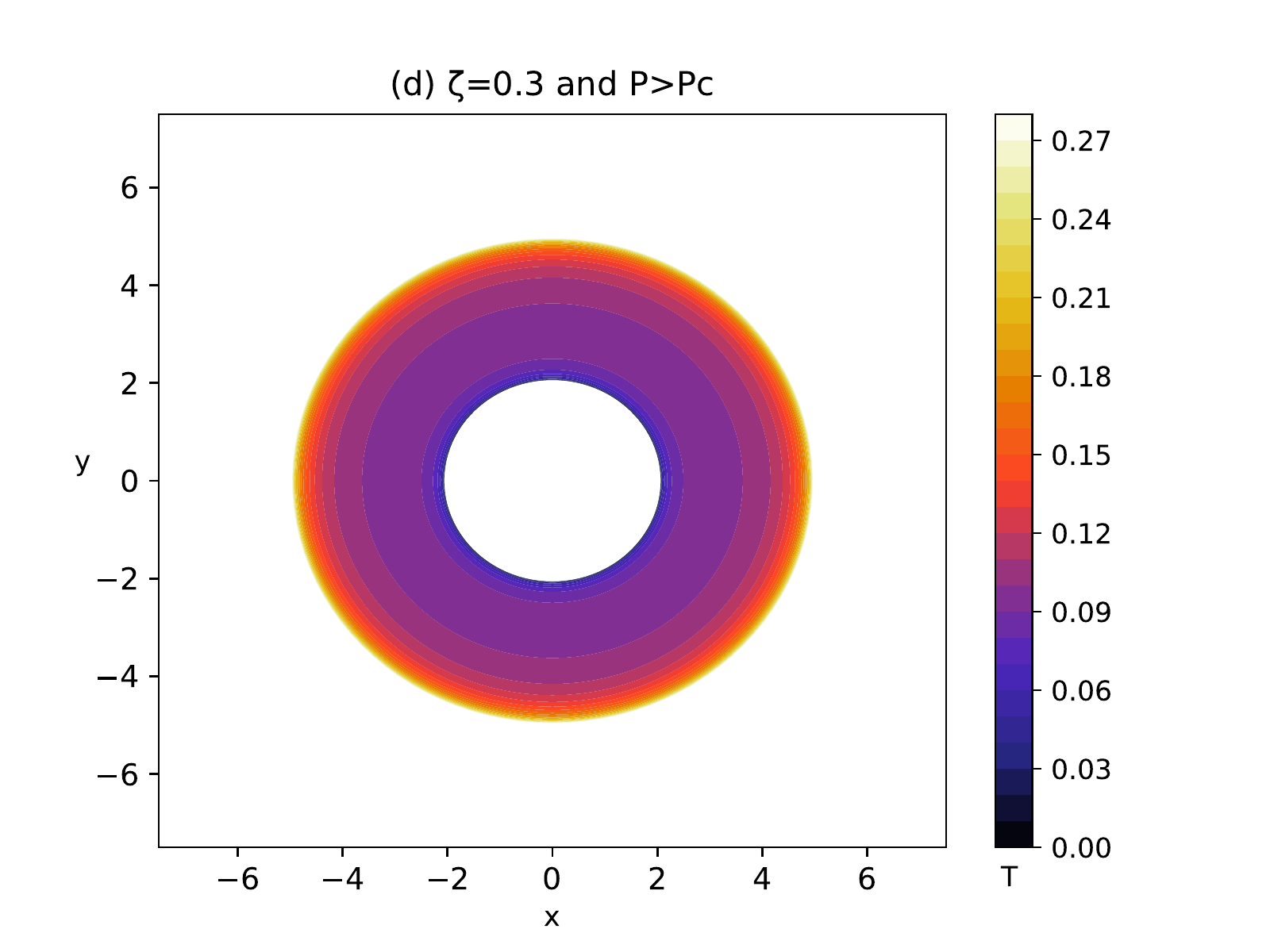}
	\includegraphics[width=4.5cm,height=4cm]{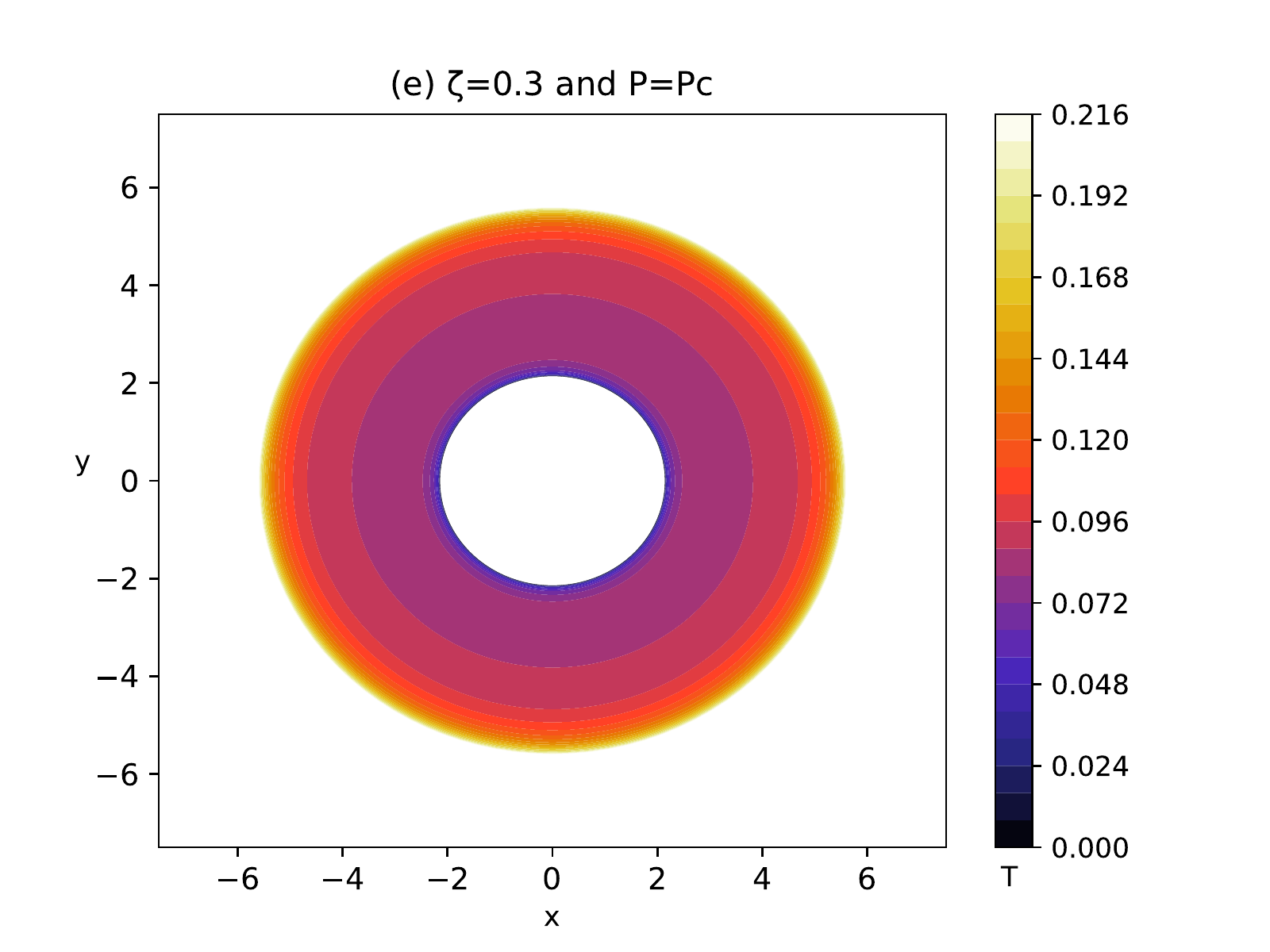}
	\includegraphics[width=4.5cm,height=4cm]{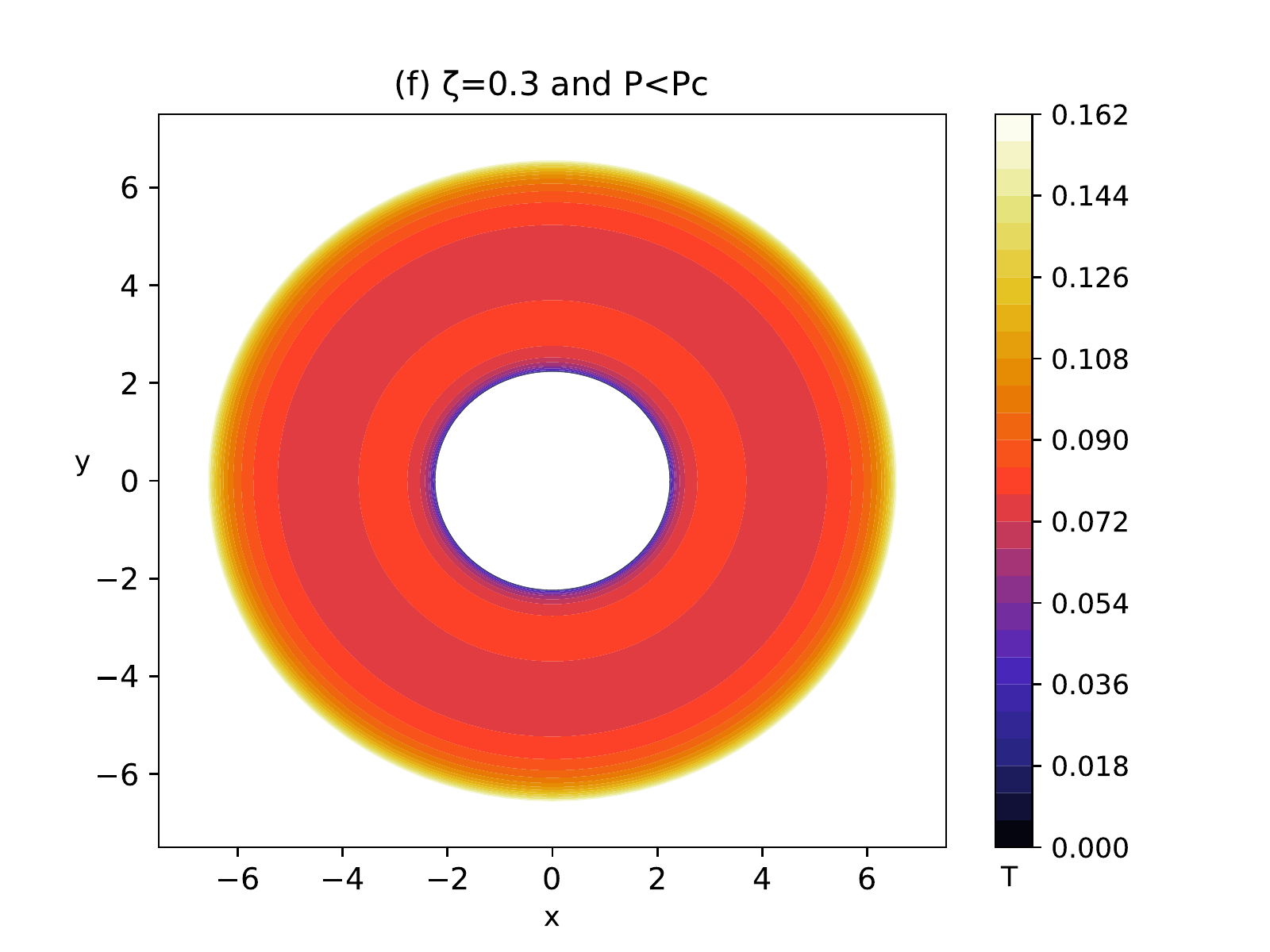}
	\parbox[c]{15.0cm}{\footnotesize{\bf Fig~5.}
		Thermal profile of the Reissner-Nordström-ADS black hole with the effects of NLED for different thermodynamical case. {\em Panel (a)}--$P>P_{\rm c}$ with $\zeta=0.1$, {\em Panel (b)}--$P=P_{\rm c}$ with $\zeta=0.1$ and {\em Panel (c)}--$P<P_{\rm c}$ with $\zeta=0.1$, {\em Panel (d)}--$P>P_{\rm c}$ with $\zeta=0.3$, {\em Panel (e)}--$P=P_{\rm c}$ with $\zeta=0.3$ and {\em Panel (f)}--$P<P_{\rm c}$ with $\zeta=0.3$. The BH mass is taken as $M=60$.}
	\label{fig5}
\end{center}
To see the ``N-type'' change more clearly, we focus on the range of $r_{\rm s}$ to ($r_{\rm s1}$ $\rightarrow$ $r_{\rm s2}$).
The corresponding results obtained are presented in Fig.6.
Finally, it turns out that the effects of the coupling constant on the Phase transition of a charged AdS black hole can also be clearly seen in the shadow context from the Fig.6, which is coincide with that obtained in Fig.2.
\begin{center}
	\includegraphics[width=5.5cm,height=4cm]{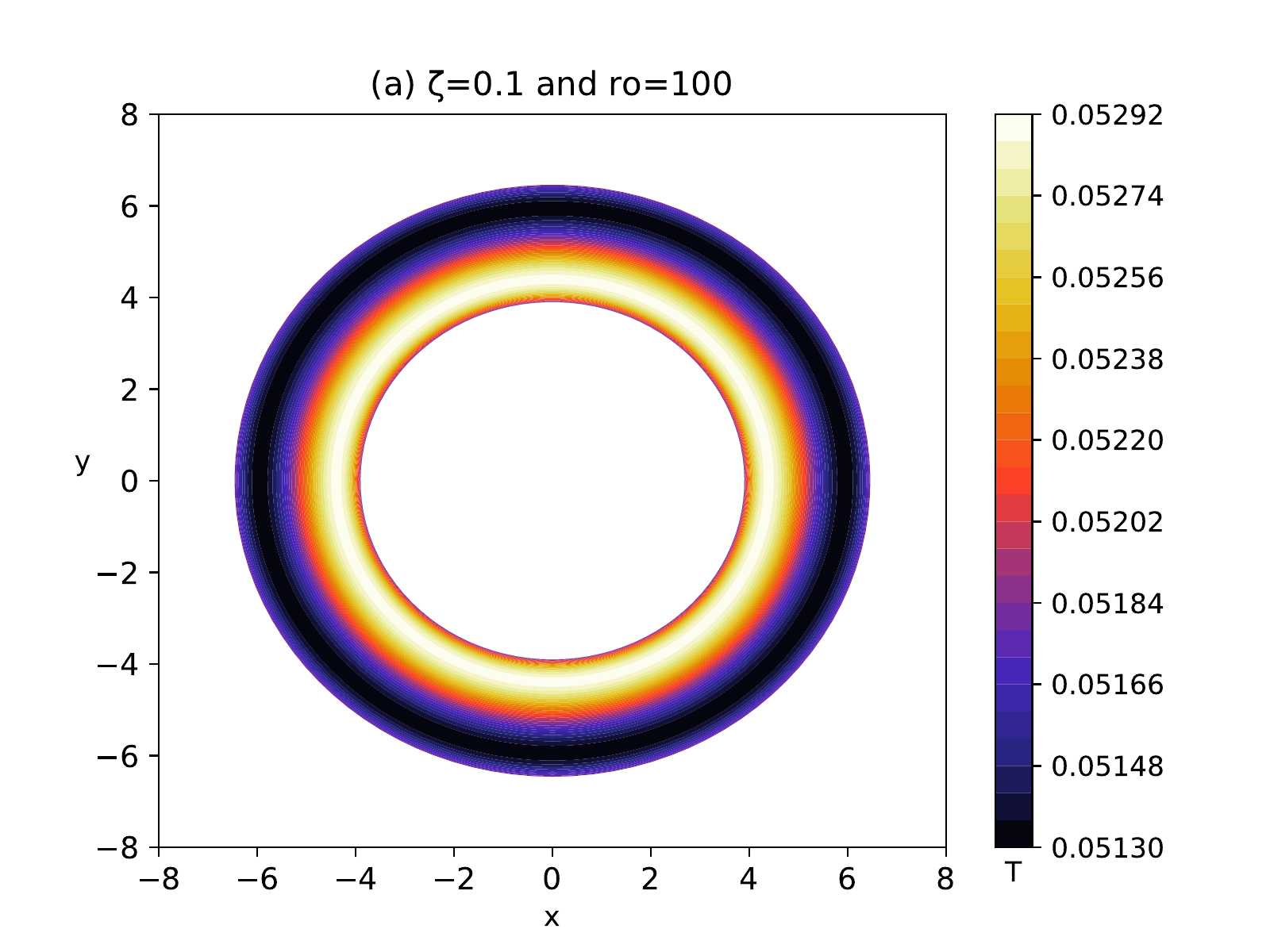}
	\includegraphics[width=5.5cm,height=4cm]{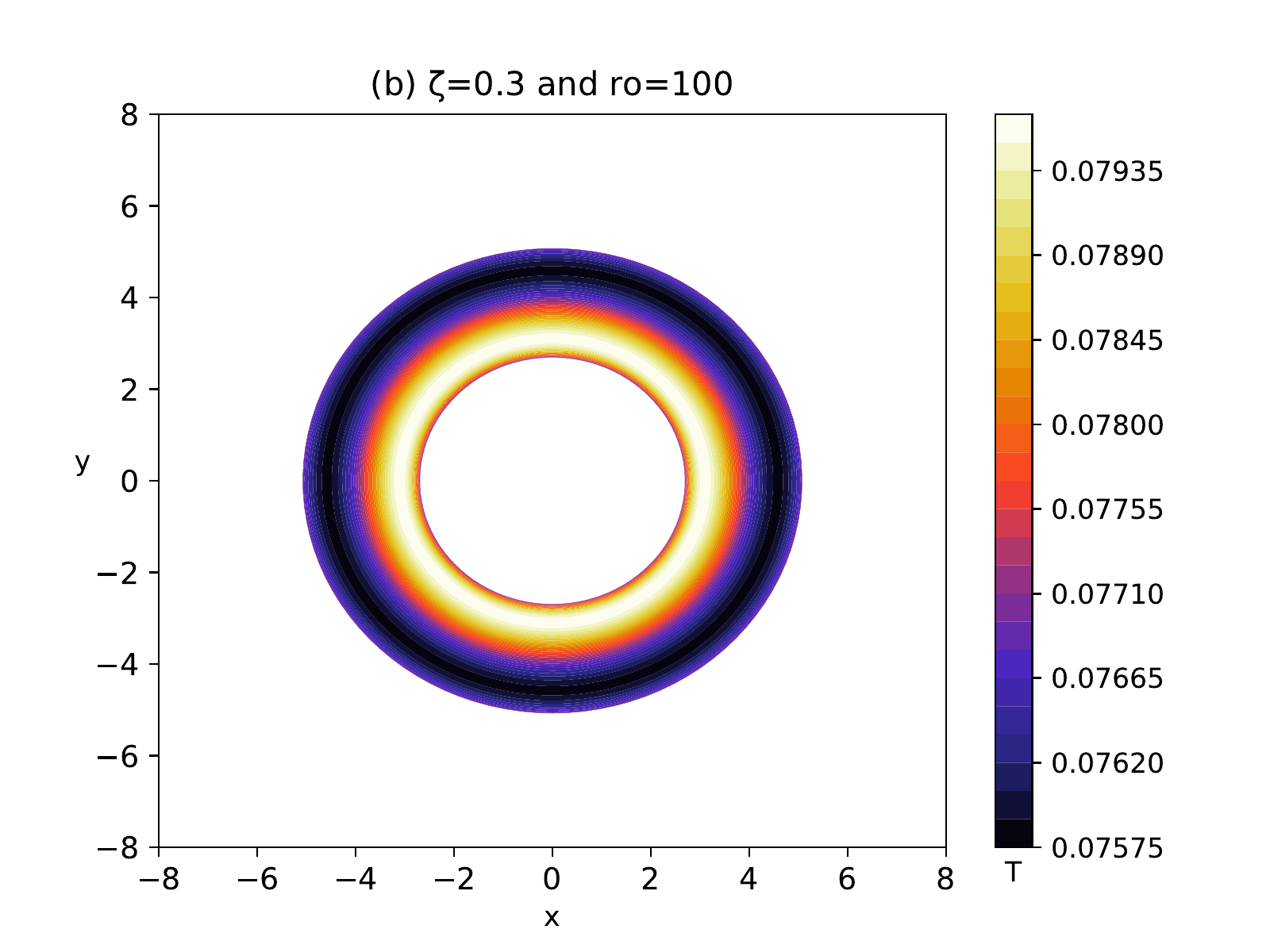}
	\includegraphics[width=5.5cm,height=4cm]{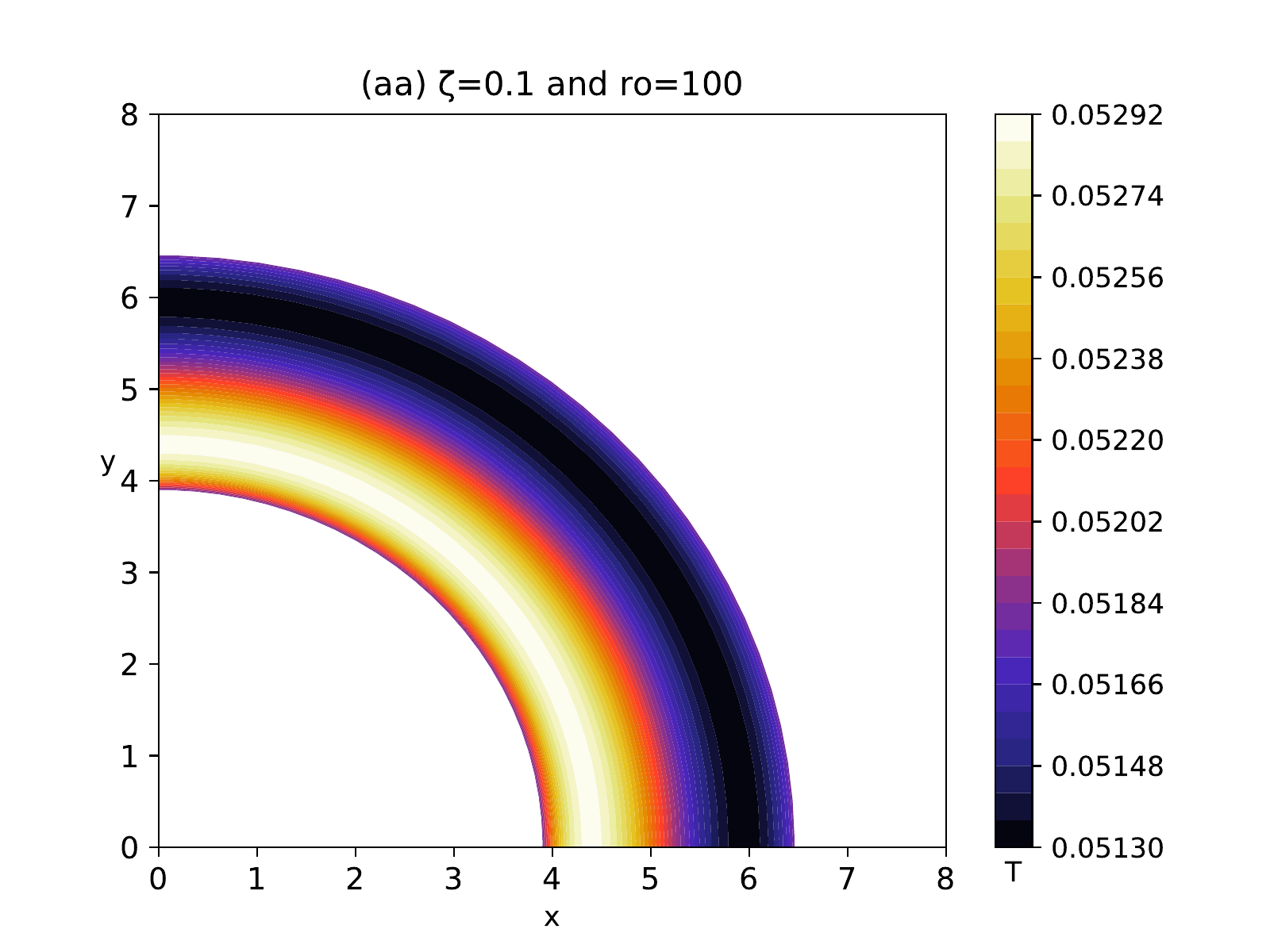}
	\includegraphics[width=5.5cm,height=4cm]{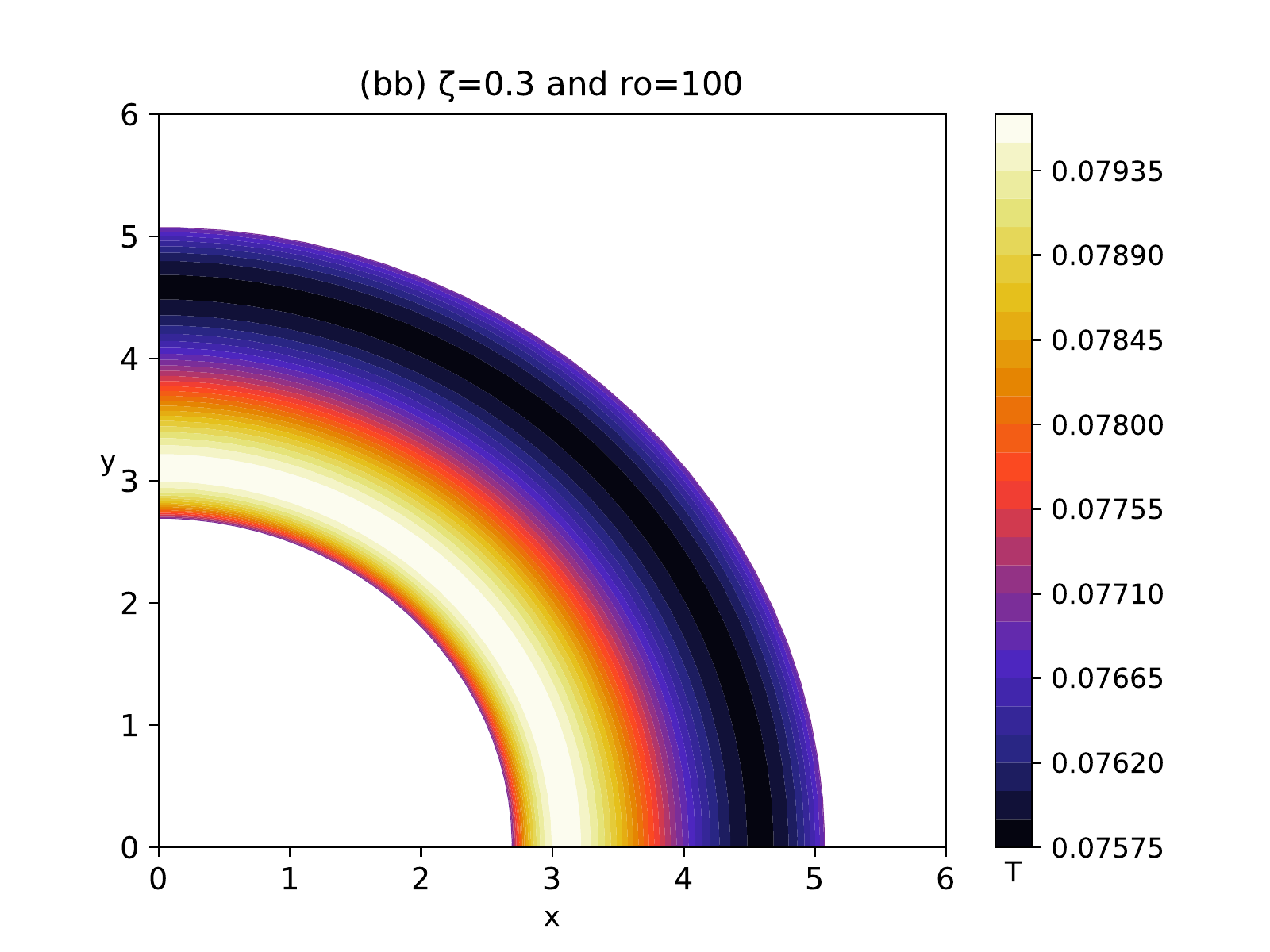}
	\parbox[c]{15.0cm}{\footnotesize{\bf Fig~6.}
		Thermal profile of a charge AdS black hole with NLED term for different thermodynamical case. {\em Panel (a)}--$P<P_{\rm c}$ with $\zeta=0.1$, {\em Panel (b)}--$P<P_{\rm c}$ with $\zeta=0.3$ and {\em Panel (aa)}--$P<P_{\rm c}$ with $\zeta=0.1$, {\em Panel (bb)}--$P<P_{\rm c}$ with $\zeta=0.3$. {\em Panel (aa)} is a quarter of {\em Panel (a)}, {\em Panel (bb)} is a quarter of {\em Panel (b)}. The black hole
 mass is taken as $M=60$.}
	\label{fig6}
\end{center}

\section{ Conclusions and discussions} \label{Conclusions}

In this paper, we have used the shadow radius instead of the horizon radius to analyze the phase transition of a charged AdS black hole with the NLED term.
First, the relationship between the shadow radius $r_s$ and event horizon radius $r_h$ is established with the aid of the corrected effective geometry which caused by the NLED term.
The results shows that they have a positive correlation, which implies that black hole temperature can also be established with the shadow radius.
When the coupling constant $\zeta \rightarrow 0$, the expression (\ref{eq: 2.32}) can also be used to describe the relationship between the shadow radius and the event horizon in the RN-AdS black hole.
In Fig.1, we present this relationship, and find that the event horizon radius $r_h$ increased, the shadow radius $r_s$ will increase. And, the trend of $r_s$ will be gradually flat with increase of $r_h$.
In view of this, we believed that the shadow can be used to present the phase transition structure of black hole thermodynamics.

Then, we further constructed the $T-r_{\rm s}$ function on the basis of the $T-r_{\rm h}$ function, and carefully analyse the phase transition curve of the AdS black hole.
In the supercritical state($P>P_c$), the $T-r_{\rm s}$ curve has no inflection point.
In a critical state($P=P_c$), black holes are in an unstable state.
Below the critical pressure($P<P_c$), there exhibits a two-phase transition branch of black hole.
When the shadow radius is small ($r_{\rm s} < r_{\rm s1}$), it corresponds to a stable small black hole. And, the case ($r_{\rm s} > r_{\rm s2}$) corresponds to a stable large black hole.
While the unstable intermediate branch appears in the range ($r_{\rm s1} < r_{\rm s} < r_{\rm s2}$).
All of those features is well coincide with that obtained by using $r_h$.
Then, combining the relationship between the heat capacity and the event horizon radius, the relationship between the heat capacity and the shadow radius is established to address phase transition grade, which are shown in Fig.3.
One can see that the heat capacity diverges at the critical point, which implies the emergence of the second-order phase transition of black hole, which is also consistent with that given by $r_h$.
Finally, based on the function $T-r_{\rm s}$, the thermal profiles of the black holes in several representative sets of coupling constants are established.
We find that the shadow radius is closely related to the pressure.
As the pressure increases, the shadow radius will decrease, but the black hole temperature will increase.
When ($P<P_{\rm c}$), it shows that the temperature variation law of the black hole is presented as ``increasing $\rightarrow$ decreasing $\rightarrow$ increasing (N-type)''. That is to say, the thermodynamics of black holes can be completely reflected on the thermal profile, and the black hole shadow can also reflect the thermodynamic phase transition relationship of black holes in a charged AdS black hole with the NLED term. In addition, the effects of the coupling coefficient $\zeta$ on thermodynamic phase transition have also been discussed throughout of paper.
Combined with above facts, we can conclude that the shadow can indeed replace the event horizon, which opens a new door for the study of thermodynamic phase transition of black hole. Finally, the effects of NLED have been carefully analysed through out the paper.

\section*{Acknowledgements}
This work is supported by the National Natural Science Foundation of China (Grant No.11903025), and by the starting found of China West Normal University (Grant No.18Q062), and by the Sichuan Youth Science and Technology Innovation Research Team (21CXTD0038), and by the Chongqing Science and Technology Bureau (csts2022ycjh-bgzxm0161), and by the Natural Science Foundation of Sichuan Province (2022NSFSC1833).


\end{document}